\documentclass[
twoside,
twocolumn,
superscriptaddress,
nofootinbib,
amsmath,
amssymb,
aps,
prd,
longbibliography
]{revtex4-2}
\usepackage{xcolor}
\usepackage{textcomp}
\usepackage{amsmath}
\usepackage{amsthm}
\usepackage{amstext}
\usepackage{amssymb}
\usepackage{stmaryrd}
\usepackage{graphicx}
\usepackage{float}  
\usepackage{tikz-feynman}
\usepackage{nicefrac}
\usepackage{multirow}
\usepackage{slashed}
\usepackage{dsfont}
\usepackage{lipsum}

\usepackage[colorlinks=true,linkcolor=magenta,citecolor=magenta,urlcolor=magenta]{hyperref}

\usetikzlibrary{patterns,decorations.markings}
\tikzset{
cross/.style={fill=white,path picture={\draw[black] (path picture bounding box.south east) -- (path picture bounding box.north west) (path picture bounding box.south west) -- (path picture bounding box.north east);}},
    dressed/.style={fill=white,postaction={pattern=north east lines}},
    momentum/.style 2 args={->,semithick,yshift=5pt,shorten >=5pt,shorten <=5pt},
    loop/.style 2 args={thick,decoration={markings,mark=at position {#1} with {\arrow{>},\node[anchor=\pgfdecoratedangle-90,font=\footnotesize] {$p_{#2}$};}},postaction={decorate}},
    label/.style={thin,gray,shorten <=-1.5ex}
}
\usetikzlibrary{patterns}
\usepackage{xcolor}
\usetikzlibrary{patterns,decorations.markings}
\def\radius{1.4}

\definecolor{mycyan}{RGB}{0, 255, 255}     
\definecolor{mymagenta}{RGB}{255, 0, 255}  
\definecolor{myteal}{RGB}{0, 128, 128}
\definecolor{vibrantteal}{HTML}{00FFCC}
\definecolor{azureblue}{HTML}{007FFF}

\makeatletter
\def\tagform@#1{(\textcolor{magenta}{#1})}
\makeatother
\usepackage[colorlinks=true,linkcolor=magenta,citecolor=magenta,urlcolor=magenta]{hyperref}%

\begin{document}

\title{\textcolor{teal}{\bf Functional Dimensional Regularization}}

\author{P. Beretta}
\email{pberetta@fing.edu.uy}
\affiliation{IFFI, Universidad de la Rep\'ublica, J.H.y Reissig 565, 11300 Montevideo, Uruguay}

\author{A. Codello}
\email{alessandro.codello@unive.it}
\affiliation{DSMN, Ca' Foscari University of Venice, Via Torino 155, 30172 - Venice, Italy}
\affiliation{IFFI, Universidad de la Rep\'ublica, J.H.y Reissig 565, 11300 Montevideo, Uruguay}

\begin{abstract}
We introduce and develop Functional Dimensional Regularization (FDR), a novel functional renormalization group (RG) framework that extends the principles of dimensional regularization beyond the perturbative $\varepsilon$-expansion. The key insight is to define the FDR beta functions, valid in any continuous dimension $d$ and characterized by threshold functions, as a \textit{sum over all critical dimensions} of the corresponding scheme-independent beta functions computed in DR. We demonstrate that this formalism exhibits all the hallmarks of a fully-fledged functional RG. Within the local potential approximation, we perform a general fixed point analysis, recovering the $\varepsilon$-expansion for all multi-critical models and establishing a direct connection to functional scaling solutions, spike plots, and eigen-perturbation spectra. Extending the framework to the second order of the derivative expansion, we compute the critical exponents for the \texttt{Ising} universality class in $d=3$, providing estimates for the anomalous dimension $\eta$ and RG spectrum that converge rapidly and show favorable agreement with state-of-the-art results. We present a detailed comparison with non-perturbative and proper-time RG approaches, highlighting structural similarities and key differences. In particular, we introduce a ``distillation'' procedure that allows one to systematically derive the FDR flow from any other one-loop-exact RG scheme. Our results establish FDR as a self-contained and competitive RG framework that combines the technical simplicity of dimensional regularization with the versatility of the functional approach.
\end{abstract}

\date{\today}

\maketitle

\section*{Introduction}\vspace{-0.2cm}

Dimensional regularization (DR) \cite{Bollini:1972ui,tHooft:1972tcz,tHooft:1973mfk,Cicuta:1972jf,Cicuta:1972zi} has been the main tool for renormalization group (RG) calculations since the early days of the field-theory approach to critical phenomena, especially when combined with the $\varepsilon$-expansion \cite{Wilson:1971dc,Wilson:1971vs,Wilson:1973jj}. In this framework one studies a fixed point just below its upper critical dimension $d_c$ in powers of $\varepsilon=d_c-d$. This combination rapidly became the standard approach in the field, and the common lore since then has been that the price to pay for using DR -- its simplicity, minimal scheme dependence, and symmetry-preserving properties -- is that it works only perturbatively around $d_c$.
In a recent paper \cite{Beretta:2026zcy}, the authors showed that it is instead possible to overcome this fundamental limitation. One can extend DR beyond the $\varepsilon$-expansion and define a well-behaved and efficient functional RG flow -- termed \textit{functional dimensional regularization} (FDR) -- that is valid in any dimension and that furthermore does not rely on a hard mass regulator. The construction is based solely on the principles of DR, yet it yields a genuinely \textit{functional} framework.

The key insight of the FDR approach is that (one-loop) beta functions -- or more generally beta functionals -- applicable in an arbitrary continuous dimension $d$, can be expressed as a \textit{sum over all critical dimensions} of the corresponding perturbative DR contributions:
\begin{equation}\label{masterformula}
\beta^{\rm FDR}(d)=\sum_{d_c}\mu^{d-d_c}\,\beta^{\rm DR}(d_c)\,.
\end{equation}
The range of the sum over critical dimensions defines the specific variant of the FDR formalism one adopts and will be discussed in detail later. Crucially, the sum must run over \textit{all} critical dimensions, including the infrared ones -- those that, from the traditional four-dimensional perspective, do not give rise to UV divergences. It is precisely the inclusion of these IR critical dimensions that allows FDR to capture threshold effects and IR physics and to correctly describe strongly coupled fixed points in low dimensions, overcoming the limitations of the conventional $\varepsilon$-expansion.

The goal of the present work is to demonstrate that the FDR formalism possesses all the distinctive features of a functional RG framework. Specifically, we show its direct applicability in $d=3$ without recourse to any expansion parameter; the emergence of threshold behavior derived from the sum over critical dimensions; the existence of functional scaling solutions that evolve continuously with the dimension; a well-posed eigen-perturbation problem for the RG spectrum; the ability to reproduce the results of the $\varepsilon$-expansion at leading order and beyond; and, finally, a meaningful and systematically improvable notion of derivative expansion. Taken together, these results establish FDR as a competitive and self-contained functional RG scheme.

The paper is organized as follows. We start with a derivation of the master formula \eqref{masterformula} and then introduce the derivative expansion in the FDR context. As a first application, we study the local potential approximation (LPA): we perform a general fixed point analysis by solving the system of beta functions in a field expansion, then recover the $\varepsilon$-expansion around all the critical dimensions related to multi-critical models, and finally study scaling solutions and eigen-perturbations directly in $d=3$. Subsequently, we set up the derivative expansion to second order (DE2) and, after performing the necessary consistency checks, study the \texttt{Ising} universality class in three dimensions, discussing the quality of different approximations to the DE2 RG flow. We conclude with a direct comparison of FDR with the non-perturbative RG based on the Wetterich--Morris equation, as well as with the proper-time RG, highlighting both similarities and structural differences. Finally, we introduce the \textit{distillation} procedure, which provides a systematic way to obtain the FDR flow from any one-loop-exact RG flow.

\section*{Functional Dimensional Regularization}\label{derivation}\vspace{-0.2cm}

In our previous paper \cite{Beretta:2026zcy}, we introduced the one-loop master formula \eqref{masterformula} in two complementary ways: 1) as a means of encoding into exponential threshold functions all the universal, scheme-independent information contained in DR beta functions; and 2) as being equivalent to standard DR with the fundamental twist of removing all $1/\varepsilon$-poles that appear in perturbation theory, not merely those associated with the specific dimension under consideration. In this section, we present a third derivation of the one-loop master formula by constructing the all-dimensions counter-term action. We then offer some comments on how \eqref{masterformula} might generalize beyond one-loop, review the three strategies outlined in \cite{Beretta:2026zcy}, and finally explain how the derivative expansion fits into the FDR formalism.

\subsection*{\color{teal}One-loop Master Formula}\vspace{-0.2cm}

It is a well-known fact in statistical physics and quantum field theory (see any standard textbook, e.g.~\cite{Zinn-Justin:2002ecy}) that quantities computed from the bare action $S$ -- which contains the bare, dimension-full coupling constants $g_i$ -- suffer from UV divergences.
A prototypical example is the one-loop effective action
\begin{equation}\label{oneloop}
\Gamma_1[\phi] =\frac{1}{2}{\rm Tr} \log S^{(2)}[\phi] \,.
\end{equation}
To fix this problem, one adopts a renormalization procedure: one writes the bare parameters in terms of the renormalized couplings $g_i^R$ and counter-terms, with the counter-terms chosen to absorb the UV divergences.
A difference between the bare  $g_i$ and renormalized $g^R_i$ couplings is that the former do not depend on the renormalization scale $\mu$, that is inevitably introduced by the regularization process, even in the case of a massless scheme such as DR.
The bare action can be expressed as a sum of terms in which a coupling constant multiplies the corresponding operator,
$S = \sum_i g_i \, \mathcal{O}^i$, and similarly the renormalized action is defined as $S_R = \sum_i g_i^R \, \mathcal{O}^i$.
Here we use the notation $\mathcal{O}^i$ to denote a basis of integrated operators, i.e. functionals of the fields, which we take to be complete and orthogonal.
In the case of the single-component scalar theory considered in this work, an example is the family of local operators without
derivatives, $\int\!\frac{1}{n!} \phi^n$.
The bare action relates with the renormalized action as
\begin{equation}\label{bareandsr}
S=S_R + \Delta S_1+...\,, 
\end{equation}
where $\Delta S_1$ is the one-loop counter-term action that has to be fixed in order to cancel the divergences of $\Gamma_1$.
In DR we compute the effective action \eqref{oneloop} for an arbitrary (complex) value of the dimension $d$ and  generally find a discrete set of critical dimensions (the $d_c$'s) around which it is divergent $\Gamma_1 \sim \tfrac{1}{\varepsilon}$, where $\varepsilon=d_c-d$.  
A key feature of DR is that -- at one-loop -- the $1/\varepsilon$-poles are directly related to the RG beta functions \cite{tHooft:1973mfk}.
Thus the pole part of the effective action at each $d_c$ has the form
\begin{equation}\label{divpart}
\Gamma_{1}|_{d_c,\infty}=\frac{\mu^{d-d_c}}{d-d_c} \sum_i \beta^{\rm DR}_i(d_c)\, \mathcal{O}^i\,,
\end{equation}
where  $\mu^{d-d_c}$ is the usual factor to keep the correct dimensionality.
In a bit more fancy words, the residues of $\Gamma_1$ are the one-loop DR beta functions. 
Now comes the key point: in the usual practice one subtracts only the pole related to the dimension of interest (for example $d_c=4$ if we are considering the $\lambda \phi^4$ theory which corresponds to the {\tt Ising} universality class, or $d_c=6$ if we are considering $g \phi^3$ and the corresponding {\tt Lee-Yang} class); in the FDR formalism instead, we subtract all the simple poles at all $d_c$'s.  Then the counter-term action is chosen as a sum over all critical dimensions:
\begin{equation}\label{DeltaS1}
\Delta S_1=-\sum_{d_c}\Gamma_{1} |_{d_c,\infty}\,.
\end{equation}
Thus, using \eqref{divpart}, we have
\begin{equation}
\Delta S_1= -\sum_{i}\sum_{d_c} \frac{\mu^{d-d_c}}{d-d_c}\,\beta^{\rm DR}_{i}(d_c)\, \mathcal{O}^{i}\,.
\end{equation}
%
Inserting this expression into \eqref{bareandsr} gives
\begin{equation}\label{barefinale}
S = S_R -\sum_{i}\sum_{d_c} \frac{\mu^{d-d_c}}{d-d_c}\,\beta_{i}^{\rm DR}(d_c)\, \mathcal{O}^{i} + ...\,.
\end{equation}
The RG flow is obtained exploiting the fact that the bare couplings $g_i$ do not depend on the renormalization scale $\mu$.
Thus their scale derivative is zero, as is the derivative of the bare action $\partial_\mu S =0$.
Deriving  both sides of \eqref{barefinale}  gives as a result
\begin{equation*}
\mu \partial_\mu S=\mu \partial_\mu S_R -\mu \partial_\mu \Biggl(\sum_{d_c} \frac{\mu^{d-d_c}}{d-d_c}\sum_{i}\beta^{\rm DR}_{i}(d_c)\, \mathcal{O}^{i}\Biggr)  \,.
\end{equation*}
Now the scale derivative of $\mu^{d-d_c}$ exactly cancels the respective simple pole $1/(d-d_c)$.
We discard here the derivative of the term $\beta^{\rm DR}(d_c)$ which gives a higher order contribution.
We now collect terms and define the FDR beta functions $\beta^{\rm FDR}_{i}(d) \equiv \mu \partial_\mu g_i^R$ to obtain:
\begin{equation}
0= \sum_{i}\Biggl(\beta^{\rm FDR}_{i}(d) -\sum_{d_c} \mu^{d-d_c}\beta_{i}^{\rm DR}(d_c)\Biggr)\mathcal{O}^{i}\,.
\end{equation}
By the orthogonality of our operator expansion, our final result is that the FDR beta functions for the renormalized couplings  $g_i^R$ are obtained as a sum over all critical dimensions (those that appear at one-loop) of the respective, traditional, DR beta functions:
\begin{equation}\label{masterformulabis}
\beta_i^{\rm FDR}(d) = \sum_{d_c} \mu^{d-d_c} \beta_i^{\rm DR} (d_c)\,.
\end{equation}
This is our (one-loop) master formula and the main result of our approach.
In conclusion, the (one-loop) FDR  beta functions are the beta functions obtained in the standard DR procedure, with the only twist that all poles are subtracted in the definition of the counter-term action. 

In \cite{Beretta:2026zcy} we proposed three options to go beyond the one-loop master formula \eqref{masterformulabis}.
Option (a) retains only the leading order critical dimensions at each loop order, which preserves the full universality of the FDR beta functions by excluding any scheme-dependent effects.
Option (b) also includes the critical dimensions beyond leading order, introducing scheme-dependent terms and thereby sacrificing strict universality for potentially improved quantitative accuracy.
Option (c) adopts a one-loop RG improvement point of view: one keeps only one-loop contributions but includes all possible operators in the action, as is typically done in functional RG.
In this paper we follow option (c), carry out a complete exploration by performing a derivative expansion to second order, and leave the exploration of options (a) and (b) to future work.
\begin{figure*}
\centering
\includegraphics[width=0.60\columnwidth]{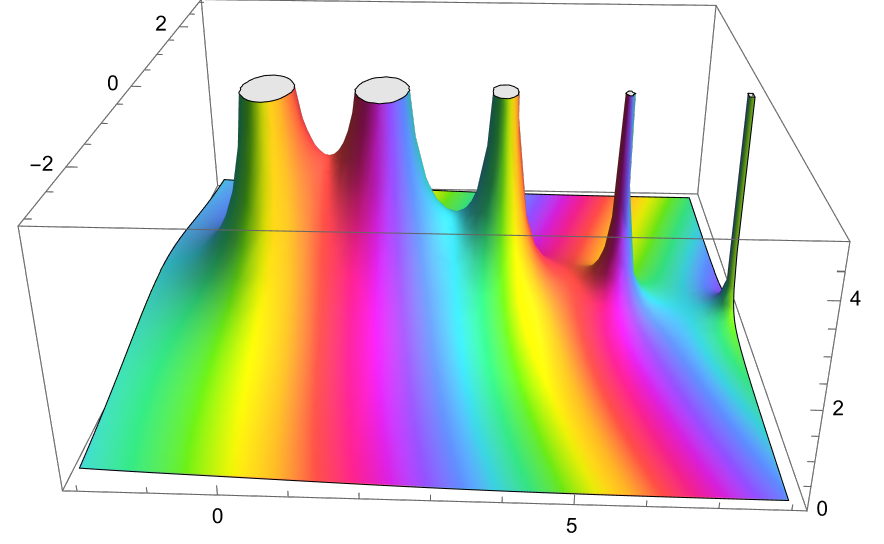}\quad
\includegraphics[width=0.60\columnwidth]{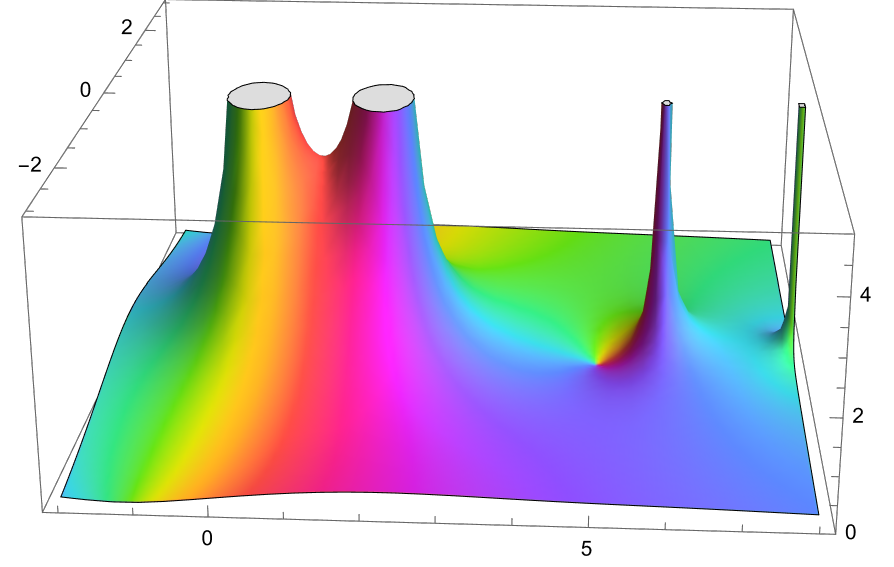}\quad
\includegraphics[width=0.60\columnwidth]{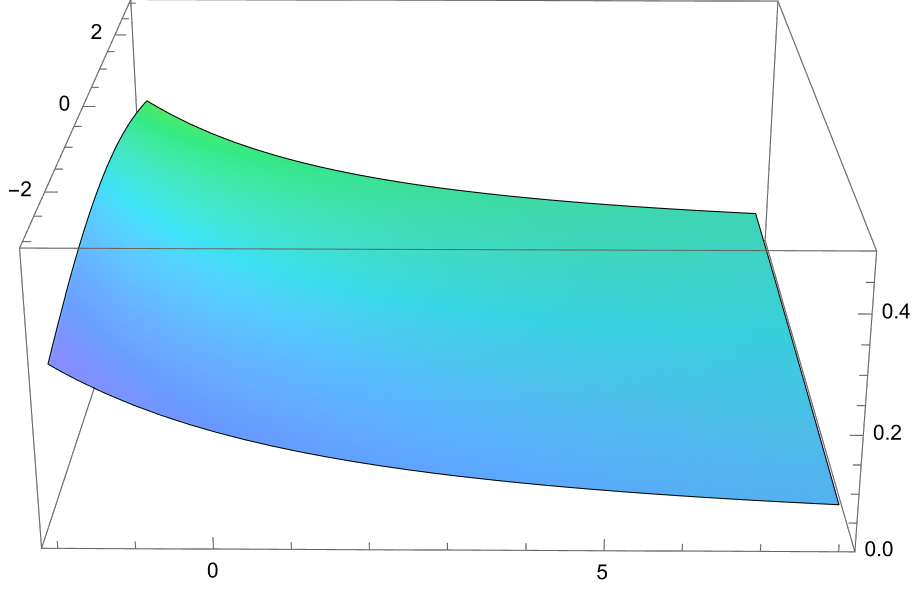}%
\caption{Left: Poles of the Gamma function $\Gamma(-\tfrac{d}{2}) \sim \Gamma_1(d)$ in the complex-$d$ plane. Middle: Poles of the Gamma function with the $d_c=4$ pole subtracted. Right: Gamma function in the complex plane after all poles have been subtracted.}
\label{poles}
\end{figure*}

\subsection*{\color{teal}Derivative Expansion}\vspace{-0.2cm}

In this paper we will explore the one-loop RG improvement point of view in the FDR context: we will consider the most general action $S$, compatible with the symmetries and degrees of freedom of the problem, and determine the RG flow it generates. 
As in similar studies with other functional RGs, the action is typically expressed as a derivative expansion (DE), where the action is expanded in powers of the derivatives/momenta with running functions encoding any possible field dependence: 
\begin{equation}\label{DEaction}
S=\int {\rm d}^{d}x \Bigl\{ \frac{1}{2}(\partial \phi)^2+ V(\phi)+\frac{1}{2}Z(\phi)(\partial \phi)^2+... \Bigr\}
\end{equation}
The relative beta functionals for $V$ and $Z$ in DR at $d_c$ are directly related to the poles as follows,
\begin{equation}\label{DEdiv}
	\Gamma_{1}|_{d_c,\infty}=-\frac{1}{\varepsilon}\int {\rm d}^{d_c}x\, \Bigl\{ \beta^{\rm DR}_{V}(d_c) +\beta^{\rm DR}_Z(d_c) \frac{1}{2}(\partial \phi)^2+\!...\! \Bigr\}\,,
\end{equation}
where $\varepsilon=d_c-d$. 
The master formula that defines the FDR flow of the potential in any dimension $d$ is:

\begin{equation}\label{master}
\beta_V^{\rm FDR}(d) \equiv \sum_{d_c} \mu^{d-d_c} \beta_V^{\rm DR} (d_c)\,,
\end{equation}
where the sum covers all possible upper critical dimensions $d_c$.
This clearly follows from \eqref{masterformulabis} applied to the couplings of the non-derivative operators $\int \tfrac{1}{n!}\phi^n$, if we see the potential as being expanded in this basis.
Similarly, for the wave-function renormalization we have

\begin{equation}\label{masterZ}
\beta_Z^{\rm FDR}(d) \equiv \sum_{d_c} \mu^{d-d_c} \beta_Z^{\rm DR} (d_c)\,,
\end{equation}
where now we are considering a two-derivative operator basis $\int \tfrac{1}{n!}\phi^n \tfrac{1}{2}(\partial \phi)^2$.
The main novel contribution of this paper -- after a review of FDR-LPA -- will be the derivation of the FDR-DE2 flow.

Before deriving the beta functions from the beta functionals, we need to rewrite these in terms of dimensionless variables
\vspace{-0.2cm}
\begin{eqnarray}\label{dimlessV}
v(\varphi) &=& \mu^{-d} V(\varphi \mu^{\frac{1}{2}(d-2+\eta)}) \nonumber\\
1+z(\varphi) &=& \mu^{\eta}\left[1+Z(\varphi\mu^{\frac{1}{2}(d-2+\eta)})\right]\,,
\end{eqnarray}
where $\mu$ is the renormalization scale, $\varphi$ the dimensionless field and $\eta$ is the anomalous dimension.
The dimensionless beta functionals $\beta_v \equiv \mu \partial_\mu v$ and $\beta_z \equiv \mu \partial_\mu z$ are easily obtained by deriving \eqref{dimlessV}:
\begin{eqnarray}
\beta_v &=&
-d \,v + \tfrac{d-2+\eta}{2}\varphi \,v' + \mu^{-d}\beta_V   \label{dimlessbetav}
\\
\beta_z &=&
\eta \, (1+z) + \tfrac{d-2+\eta}{2}\varphi\, z' + \mu^{\eta}\beta_Z \,.\label{dimlessbetaz}
\end{eqnarray}
Finally, the anomalous dimension $\eta$ is determined self-consistently by requiring that the normalization condition $z(0)=0$ be preserved by the flow, i.e.\ by imposing $\beta_z(0)=0$ in \eqref{dimlessbetaz}:
\begin{equation}\label{eta}
\eta = -\mu^{\eta}\beta_Z \Big|_{\varphi \to 0}\,.
\end{equation}
Later we will show that this last expression gives rise to very simple forms for the critical exponent $\eta$ as a function of the couplings, but which still results in quite good numerical estimates.

\section*{Local Potential Approximation\label{FPanalysis}}\vspace{-0.2cm}

The simplest functional approximation is the local potential approximation (LPA), in which only the potential fuels the RG flow while the wave-function-renormalization is frozen to its bare value, $Z=1$. In this section, after deriving the LPA flow we study the {\tt Ising} universality class with both a field expansion and a functional scaling solution, finding complete agreement. We close with a first connection with the $\varepsilon$-expansion for all multi-critical models.

\subsection*{\color{teal}Derivation}\vspace{-0.2cm}

In this section we derive the beta functional of the potential described in \cite{Beretta:2026zcy}.
We will present a simple pedagogical computation that uses the Heat Kernel formalism starting from the one-loop effective action \eqref{oneloop}.
At the LPA level the action just contains the potential in addition to the classical -- Gaussian -- kinetic term
\begin{equation}
S[\phi]=\int {\rm d}^d x \left\{ \frac{1}{2}\partial_\mu\phi\partial^\mu\phi+V(\phi) \right\}\,.
\end{equation} 
From here we can compute the Hessian $S^{(2)}$ as
\begin{eqnarray*}
S^{(2)}[\phi]=-\Box+V'' (\phi)\,,
\end{eqnarray*}
which we can insert into \eqref{oneloop} to then perform some basic Heat Kernel manipulations \cite{Vassilevich:2003xt}:
\begin{eqnarray}
\Gamma_1&=&\frac{1}{2}{\rm Tr}\log(-\Box +V'')\nonumber\\
&=&-\frac{1}{2}\int_0^\infty \frac{{\rm d}s}{s} {\rm Tr}\, e^{-s(-\Box+V'')} \nonumber\\
&=&-\Omega\frac{1}{2}\int_0^\infty \frac{{\rm d}s}{s} \frac{1}{(4\pi s)^{\frac{d}{2}}} e^{-sV''}\nonumber\\
&=&-\Omega\frac{1}{2}\frac{1}{(4\pi)^{\frac{d}{2}}}\int_0^\infty {\rm d}s \, s^{-(\frac{d}{2}+1)} e^{-sV''}\,.\label{steps}
\end{eqnarray}
We have assumed $\phi$ to be constant and we have defined $\Omega$ as the (space-time) volume.
Now we perform the proper-time integral in \eqref{steps} which gives the final expression for $\Gamma_1$, at a constant field configuration, which is proportional to an (Euler) Gamma function 
\begin{equation}\label{Gamma1}
\Gamma_1(d)= -\Omega \frac{\Gamma\big(-\frac{d}{2}\big)}{2(4\pi)^{\frac{d}{2}}} (V'')^{\frac{d}{2}}\,.
\end{equation}
The Gamma function $\Gamma(-\tfrac{d}{2})$ is the origin of the one-loop simple poles of $\Gamma_1(d)$ -- which are located at $d=2n\equiv d_c$, with $n$ natural.
To find the analytical expression for the $\frac{1}{\varepsilon}$-poles (remember $\varepsilon = d_c-d$) we just need to use the standard relation
%
\begin{equation}\label{polesGamma}
\Gamma\big(-\tfrac{d}{2}\big) = \frac{1}{\varepsilon}(-1)^{\tfrac{d_c}{2}}\frac{2}{(\tfrac{d_c}{2})!} + ...\,.
\end{equation}
Substituting this expression into \eqref{Gamma1} gives the divergent part of the one-loop action at constant field
\begin{equation}\label{potentialpoles}
\Gamma_1|_{d_c,\infty}=-\frac{1}{\varepsilon}\Omega\frac{(-1)^{\tfrac{d_c}{2}}}{(4\pi)^{\frac{d_c}{2}}(\frac{d_c}{2})!}(V'')^{\tfrac{d_c}{2}}\,.
\end{equation}
We can now extract the DR beta functionals for the potential using the relation between the betas and the poles \eqref{DEdiv}, which for constant fields boils down to $\Gamma_1|_{d_c,\infty}=-\frac{1}{\varepsilon}\Omega \,\beta^{\rm DR}_V(d_c)$.
Inserting \eqref{potentialpoles} into this last relation we find the following result
\begin{equation}\label{betaVDR}
\beta_V^{\rm DR} (d_c = 2n) =  \frac{(-1)^n}{(4 \pi)^n n!} (V'')^n\,.
\end{equation}
Plugging now this expression into the master formula \eqref{master} gives an easy summation\footnote{From now on we will drop the subscript FDR on beta functions/functionals since no confusion should arise.}
\begin{eqnarray}
\beta_V(d)  &=&  \sum_{n = 0}^\infty \mu^{d-2n} \frac{(-1)^n}{(4 \pi)^n n!} (V'')^n \nonumber\\
&=&  \mu^d \sum_{n = 0}^\infty  \frac{1}{n!} \left(\frac{-V''}{4 \pi \mu^2}\right)^n \,,
\end{eqnarray}
which returns the exponential threshold function that characterizes the FDR-LPA:
\begin{equation}\label{betaVL1}
\beta_V (d) = \mu^d \,e^{ -\tfrac{V''}{4 \pi \mu^2}}\,.
\end{equation}
Finally, by inserting \eqref{betaVL1} into \eqref{dimlessbetav} and taking into account that at the LPA level $\eta=0$ we arrive at the dimension-less FDR-LPA equation
\begin{equation}\label{flowLPAv}
\beta_v=-d\,v+\tfrac{d-2}{2}\varphi \, v'+e^{-v''}\,.
\end{equation}
In \eqref{flowLPAv} we have furthermore rescaled the field and the potential to get rid of the $4\pi$ factors.
Note that we could have, equivalently, rescaled $\mu \to \mu/\sqrt{4\pi}$.
In the continuation of the section we will study the RG flow generated by this equation.

As a final remark, the LPA approximation offers an intuitive way to see the difference between the standard one-pole subtraction procedure and the subtraction of all poles as prescribed by FDR. In Figure~\ref{poles}, the expression \eqref{Gamma1} for the one-loop effective action at constant field is plotted over the complex-$d$ plane, together with the corresponding expressions obtained by subtracting only the $d_c=4$ pole (standard procedure) and by subtracting all poles (FDR procedure). It is clear that only the latter case grants access to a renormalized effective action that is well-defined and physically meaningful in all dimensions -- a necessary requirement for any study of critical phenomena in a continuous dimension.

\subsection*{\color{teal}Field Expansion (around $\varphi=0$)}\vspace{-0.2cm}

If we expand the dimensionless potential $v(\varphi)$ in Taylor series around $\varphi=0$, then the coefficients are the dimension-less running coupling constants $\lambda_n$.
Expecting $\mathbb{Z}_2$-symmetry at the fixed point we keep only even couplings (here $n_v$ is the truncation order)
\begin{equation}\label{taylor}
v(\varphi)=\sum_{n=0}^{n_v}\frac{\lambda_{2n}}{(2n)!}\varphi^{2n} = \lambda_0+\tfrac{\lambda_2 }{2} \varphi ^2+\tfrac{\lambda _4 }{4!}\varphi ^4
+ ...\,,
\end{equation}
and their dimensionless beta functions $\beta_n \equiv \mu\, \partial_\mu \lambda_n$ can be straightforwardly extracted by inserting \eqref{taylor} into \eqref{flowLPAv} and by comparing coefficients of the same power of the field on both sides
\begin{equation}\label{betagen}
\beta_n = \frac{\partial^n}{\partial \varphi^n} \beta_v(\varphi) \Big|_{\varphi\to 0}\,.
\end{equation}
In this way we obtain the following general system of beta functions:
\begin{eqnarray}\label{betas}
\beta_{0}	& = &-d\lambda_{0}+e^{-\lambda_{2}}\nonumber\\
\beta_{2}	& = &-2\lambda_{2}-\lambda_{4}e^{-\lambda_{2}}\nonumber\\
\beta_{4}	& = &(d-4)\lambda_{4}+3\lambda_{4}^{2}e^{-\lambda_{2}}-\lambda_{6}e^{-\lambda_{2}}\nonumber\\
\beta_{6}	& = &(2d-6)\lambda_{6}-15\lambda_{4}^{3}e^{-\lambda_{2}}+15\lambda_{6}\lambda_{4}e^{-\lambda_{2}}
-\lambda_{8}e^{-\lambda_{2}}\nonumber\\
\beta_{8}	& = &(3d-8)\lambda_{8}+105\lambda_{4}^{4}e^{-\lambda_{2}}-210\lambda_{6}\lambda_{4}^{2}e^{-\lambda_{2}}\nonumber\\&&+28\lambda_{8}\lambda_{4}e^{-\lambda_{2}}+35\lambda_{6}^{2}e^{-\lambda_{2}}-\lambda_{10}e^{-\lambda_{2}}\nonumber\\
\beta_{10}& = & ...
\end{eqnarray}
The final step to determine the existence of a universality class in a given dimension $d$, is to solve for fixed points $\beta_i = 0$\,.
Since all beta functions in \eqref{betas} are linear in a coupling -- note that $\beta_n \sim \lambda_{n+2}$ -- we can iteratively solve the system in terms of $\sigma\equiv\lambda_2$ to obtain the following solution, valid near the expansion point $\varphi=0$:
\begin{equation*}
\lambda_{0}^{*}(\sigma) =  \tfrac{1}{d}e^{-\sigma}\qquad
\lambda_{2}^{*}(\sigma) =  \sigma\qquad
\lambda_{4}^{*}(\sigma) = -2e^{\sigma}\sigma
\end{equation*}
\begin{equation*}
\lambda_{6}^{*}(\sigma) =  2 e^{2\sigma}\sigma  \left\{ 6\sigma-{\color{teal}(d-4)}\right\}
\end{equation*}
\begin{equation*}
\lambda_{8}^{*}(\sigma)=  -4 e^{3\sigma}\sigma\left\{ 60\sigma^{2}+3(26-7d)\sigma+{\color{teal}(d-4)(d-3)}\right\}
\end{equation*}
\begin{eqnarray}
&&\lambda_{10}^{*}(\sigma) = 4 e^{4\sigma}\sigma\big\{ 2520\sigma^{3}+(4848-1356d)\sigma^{2}
\nonumber\\
&&+(154d^2-1074d+1856)\sigma  {\color{teal} -(d-4)(d-3)(3d-8)}\big\}\nonumber\\\label{solbetas}
\end{eqnarray}
and so on for $n_v$ arbitrarily large.
The fixed point potential can be written now in terms of $\sigma$ by substituting the exact solution for the couplings \eqref{solbetas} into the Taylor expansion \eqref{taylor}
\begin{equation}\label{taylorsigma}
v_*(\varphi)= \tfrac{1}{d}e^{-\sigma}+\tfrac{\sigma}{2}\varphi^2-\tfrac{2e^{\sigma}\sigma}{4!}\varphi^4+...\,.
\end{equation}
As in the case of the NPRG-LPA the set of solutions is parametrized by $\sigma$ and global consistency of the solution will quantize the admissible values. 
\begin{figure*}
\centering
\includegraphics[width=0.95\columnwidth]{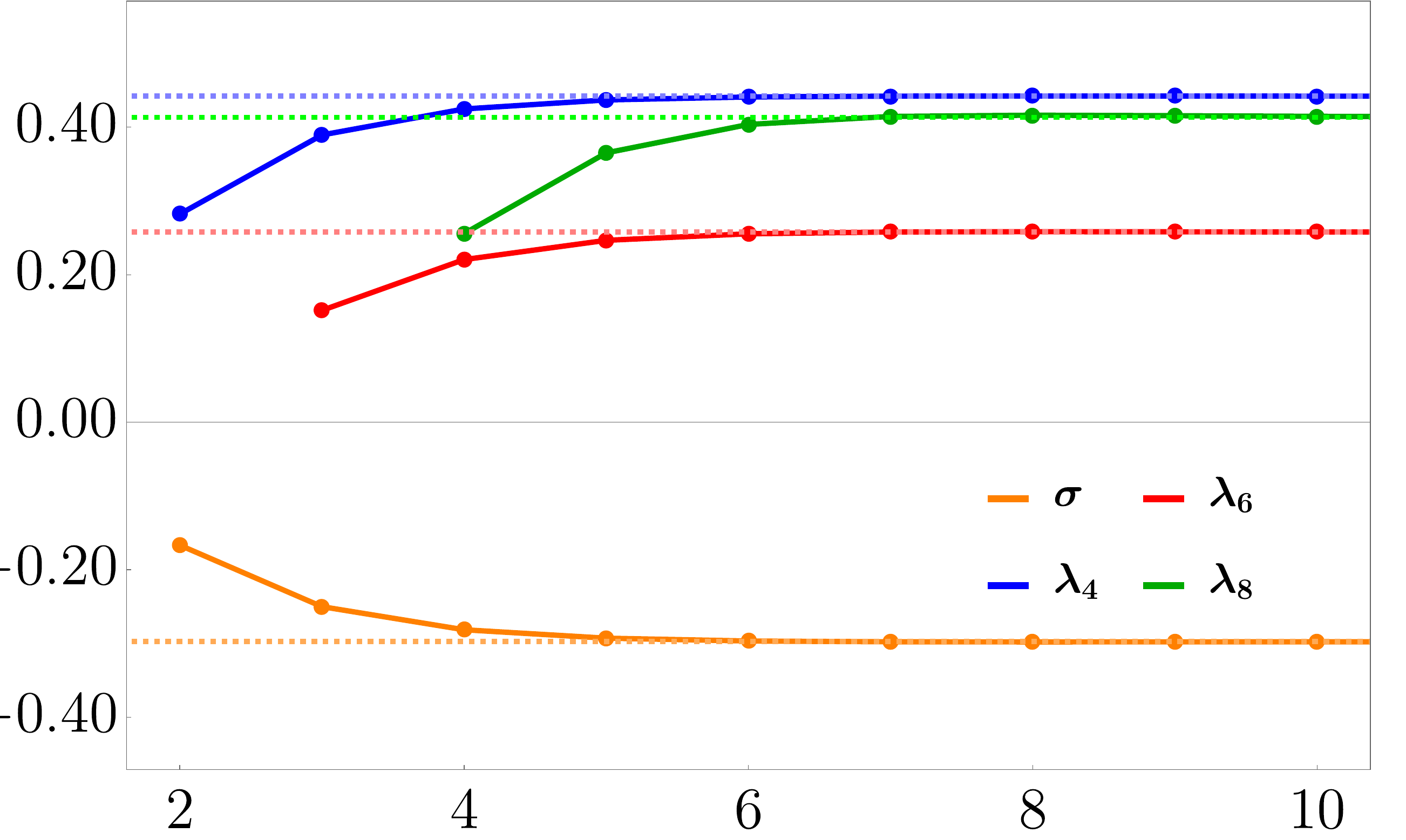}\qquad\qquad
\includegraphics[width=0.95\columnwidth]{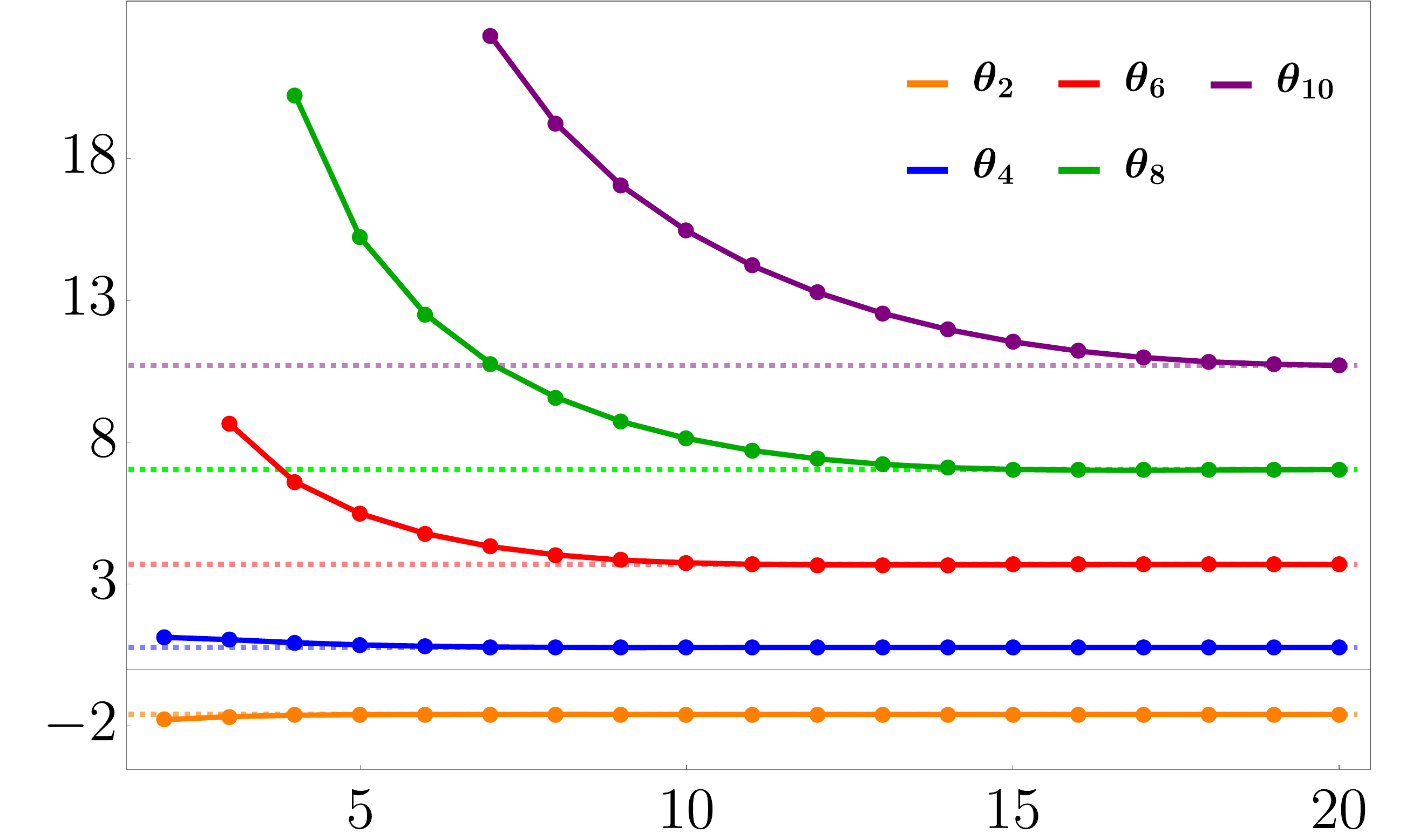}
 \caption{Wilson-Fisher fixed point values for the first couplings of the field expansion in $d=3$ (left) and eigenvalues of the stability matrix (right). Convergence is very rapid, in particular we find $\sigma_{\tt Ising} = -0.29754...$ (left).}
 \label{SpectrumLPA30smallfield}
\end{figure*}

\subsubsection*{Wilson-Fisher fixed point in $d=3$ (I)}\label{secWF1}\vspace{-0.2cm}

We now turn to the main problem our formalism is built to address: the computation of critical exponents directly in three dimensions, where the only single-field unitary universality class is {\tt Ising}.
For pedagogical reasons we go through the first polynomial truncations. In the first non-trivial case, $n_v=2$, the beta functions are the second and third of \eqref{betas} with $\lambda_6 = 0$,
\begin{equation}
\beta_2 = -2 \lambda _2- \lambda _4  e^{-\lambda _2}\,, \qquad 
\beta_4 = -\lambda_4 +3  \lambda _4^2 e^{-\lambda _2}\,.
\end{equation}
The resulting fixed point equations $\beta_2=\beta_4=0$ are transcendental but can still be solved exactly. 
The Wilson-Fisher fixed point -- representing the  {\tt Ising} universality class -- is $\lambda_2^* = -\tfrac{1}{6}$ and $\lambda_4^* = \tfrac{1}{3 \sqrt[6]{e}}$.
The stability matrix, from which we derive the RG spectrum and the associated critical exponents, is defined as usual
\begin{equation*}
M_{ij} = \frac{\partial \beta_{2i}}{\partial \lambda_{2j}}\Big|_*\,.
\end{equation*}
Evaluated at the Wilson-Fisher fixed point, it reads
\begin{equation}
\mathbb{M}=\left(
\begin{array}{cc}
 -\tfrac{5}{3} & -\sqrt[6]{e} \\[4pt]
 -\tfrac{1}{3 \sqrt[6]{e}} & 1 \\
\end{array}
\right),
\end{equation}
with eigenvalues
\begin{eqnarray}
\{\theta_2,\theta_4\} &=& \left\{\tfrac{-\sqrt[6]{e} \sqrt{19}-\sqrt[6]{e}}{3 \sqrt[6]{e}},\tfrac{\sqrt{19} \sqrt[6]{e}-\sqrt[6]{e}}{3 \sqrt[6]{e}}\right\}\nonumber\\
&=&\{-1.7863...,1.11963...\}\,.
\end{eqnarray}
The $n_v=2$ estimates for the critical exponents are thus $\nu_{n_v= 2} = -1/\theta_2 = 0.559816...$ and $\omega_{n_v = 2} = \theta_4 = 1.11963...$\,. 
At the next truncation order, $n_v=3$, the dimension-less beta functions become
\begin{eqnarray*}
\beta_2 &=&-2 \lambda _2-\lambda _4 e^{-\lambda _2}  \\
\beta_4 &=& -\lambda _4+3\lambda _4^2 e^{-\lambda _2} - \lambda _6 e^{-\lambda _2}  \\
\beta_6 &=&-15 \lambda _4^3 e^{-\lambda _2}+15 \lambda _4 \lambda _6 e^{-\lambda _2} \,,
\end{eqnarray*}
and the Wilson-Fisher fixed point sits now at
$\lambda_2^* = -\tfrac{1}{4}$, $\lambda_4^* = \tfrac{1}{2 \sqrt[4]{e}}$ and $\lambda_6^* = \tfrac{1}{4 \sqrt{e}}$.
The stability matrix takes the form
\begin{equation}
\mathbb{M}=\left(
\begin{array}{ccc}
 -\tfrac{3}{2} & -\sqrt[4]{e} & 0 \\[4pt]
 -\tfrac{1}{2 \sqrt[4]{e}} & 2 & -\sqrt[4]{e} \\[4pt]
 0 & -\tfrac{15}{2 \sqrt[4]{e}} & \frac{15}{2} \\
\end{array}
\right),
\end{equation}
whose eigenvalues are
\begin{equation}
\{\theta_2,\theta_4,\theta_6\}= \{\ -1.67498..., 1.03671..., 8.63827...\}\,,
\end{equation}
yielding $\nu_{n_v = 3} = -1/\theta_2 = 0.597023...$ and $\omega_{n_v = 3} = \theta_4 = 1.03671...$\,. 
These results are the first two points in the plots of Figure~\ref{SpectrumLPA30smallfield}.

While we can continue to higher truncation order as in these two examples, using each time the previous fixed point coordinates as numerical seeds for the successive approximation, there is a swifter procedure that hedges on the analytical solution \eqref{solbetas} found previously.
Under the truncation with $n_v$ couplings, and when relations \eqref{solbetas} are taken into account, all $\beta_i$ with $i<2n_v$ in \eqref{betas} are identically zero; only $\beta_{2n_v}$ remains unsatisfied since we have artificially set to zero $\lambda_{2n_v+2}$. It thus becomes a polynomial equation in $\sigma$ alone, and the Wilson-Fisher root $\sigma_{\tt Ising}\equiv \sigma_{n_v\to\infty}$ can be easily found with a numerical routine starting from $\sigma_{n_v=2}=-\tfrac{1}{6}$ and increasing $n_v$. 
The outcome of this full analysis is displayed in Figure~\ref{SpectrumLPA30smallfield},
where we observe a rapid convergence with $n_v$ of both the couplings $\lambda_i^*$ (obtained inserting the numerically found $\sigma_{n_v}$ into \eqref{solbetas}) and the RG eigenvalues $\theta_i$ (and hence the critical exponents).
We pushed the analysis to $n_v=20$ where full convergence is achieved for the low-lying couplings and eigenvalues.
In particular we find $\sigma_{\tt Ising} = -0.29754...$ and the following eigenvalues of the LPA spectrum:
\begin{equation}\label{thetai}
\theta_2 = -1.59750... \qquad
\theta_4 = 0.762213...
\end{equation}
\begin{equation*}
\theta_6 = 3.68532... \qquad
\theta_8 = 7.03162...\,.
\end{equation*}
From these we deduce the final FDR-LPA estimates $\nu = -1/\theta_2 = 0.6260...$ and $\omega = \theta_4 = 0.7622...$.
As remarked in \cite{Beretta:2026zcy}, the rate of convergence observed in the field expansion of the FDR-LPA is quite striking if compared to other functional RGs. Furthermore, the FDR values for both $\nu$ and $\omega$ lie closer to the Conformal Bootstrap results \cite{Chang:2024whx}, than those from the NPRG-LPA with -- optimal -- Litim regulator \cite{Litim:2001up}.

\subsubsection*{Recovering the $\varepsilon$-expansion (I)}\vspace{-0.2cm}

A general analysis can be performed for each {\tt multi-critical} universality class with $d_c=\tfrac{2n}{n-1}$ and $n=2,3,4,...$, corresponding to the Landau-Ginzburg potentials $\varphi^{2n}$. The general result computed by \cite{ODwyer:2007brp,Codello:2017hhh} in the framework of perturbation theory is
\begin{equation}\label{genspectrum}
-\theta_{i}= d-i\tfrac{d-2}{2}-2(n-1)\tfrac{n!}{(2n)!}\tfrac{i!}{(i-n)!}\,\varepsilon+O(\varepsilon^2)\,.
\end{equation}
where in the case of {\tt Ising} $n=2$ and for the {\tt Tricritical} $n=3$ as we will study below.
Multi-critical models can be already seen emerging in the highlighted factors in \eqref{solbetas}.
If $\sigma$ is of order $O(\varepsilon)$ and if we are in $d=4-\varepsilon$ then the $d-4$ factor in $\lambda_6^*$ makes it order $O(\varepsilon^2)$;
instead, if we are in $d=3-\varepsilon$ then $d-4$ will be order one and $\lambda_6^*$ becomes $O(\varepsilon)$, and so on.
Thus to recover the $\varepsilon$-expansion we fix a truncation order $n_v$ and use the last beta function of that truncation, coupled with the ansatz $\sigma = \alpha\, \varepsilon$, to univocally fix $\alpha$. Then we use \eqref{solbetas} to determine the $\varepsilon$-expansion of all the other couplings and from there we obtain the stability matrix from which we extract the RG spectrum.
\\

\paragraph*{\tt Ising}

Following this procedure in $d_{c}=4$ we find $\alpha=-\tfrac{1}{6}$ and the only non-zero fixed 
point couplings at order $O(\varepsilon)$ are
\begin{equation}\label{LPAepsilonIsing}
\lambda_2=-\tfrac{\varepsilon}{6}+O(\varepsilon^{2}) \qquad
\lambda_4=\tfrac{\varepsilon}{3}+O(\varepsilon^{2}) \,,
\end{equation}
with all other couplings at least $O(\varepsilon^{2})$.
The stability matrix is bi-diagonal
\begin{equation*}
\mathbb{M} = {\tiny\left(\begin{array}{ccccccc}
-2+\frac{\varepsilon}{3} & -1-\frac{\varepsilon}{6} & 0 & 0 & 0 & 0 & \cdots\\
0 & \varepsilon & -1-\frac{\varepsilon}{6} & 0 & 0 & 0 & \cdots\\
0 & 0 & 2+3\varepsilon & -1-\frac{\varepsilon}{6} & 0 & 0 & \cdots\\
0 & 0 & 0 & 4+\frac{19\varepsilon}{3} & -1-\frac{\varepsilon}{6} & 0 & \cdots\\
0 & 0 & 0 & 0 & 6+11\varepsilon & -1-\frac{\varepsilon}{6} & \cdots\\
0 & 0 & 0 & 0 & 0 & 8+17\varepsilon & \cdots\\
\vdots & \vdots & \vdots & \vdots & \vdots & \vdots & \ddots
\end{array}\right)}
\end{equation*}
and the first eigenvalues are
\begin{equation}
\theta_2=-2+\tfrac{\varepsilon}{3}+O(\varepsilon^{2}) \qquad \theta_4=\varepsilon+O(\varepsilon^{2})\nonumber
\end{equation}
\begin{equation}
\theta_{6}=2+3\varepsilon+O(\varepsilon^{2})\qquad\theta_8=4+\tfrac{19\varepsilon}{3}+O(\varepsilon^{2}) \nonumber
\end{equation}
\begin{equation} \label{specIsinglpa}
 \theta_{10}=6+11\varepsilon+O(\varepsilon^{2}) \qquad \theta_{12}=8+17\varepsilon+O(\varepsilon^{2}) \,.
\end{equation}
Consistently, this spectrum agrees with the one computed using perturbation theory given above \eqref{genspectrum}.
\\

\begin{figure*}
\centering
\includegraphics[width=0.95\columnwidth]{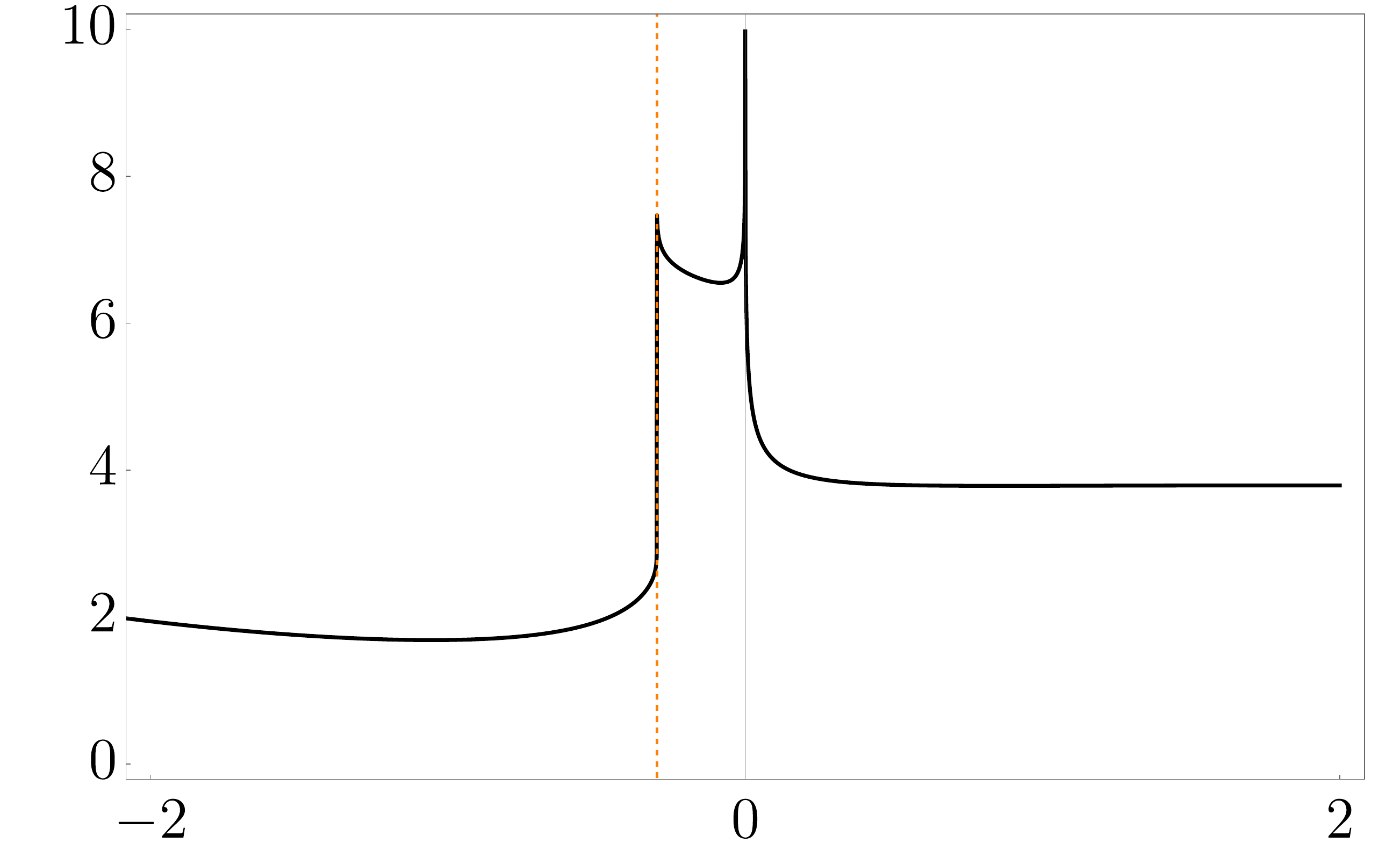}\qquad\qquad\includegraphics[width=0.95\columnwidth]{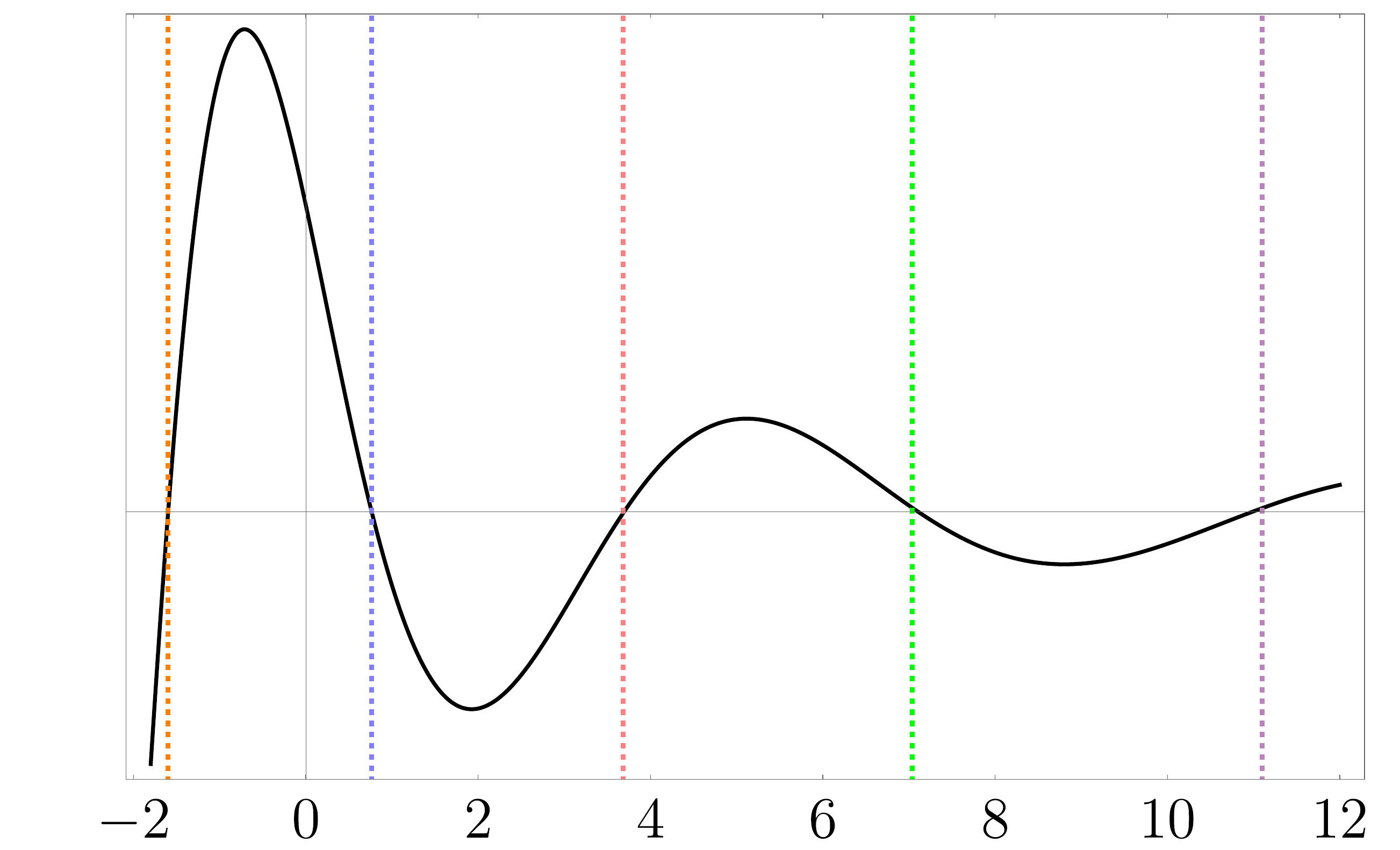}
\caption{Left: Spike plot in $d=3$. The dashed line is the fixed point value $\sigma_{\tt Ising} = -0.29754...$ found with the polynomial expansion. Right: FDR-LPA spectrum in $d=3$ obtained from the linearized LPA equation. Vertical lines are drawn respectively at $\theta = -1.59750, 0.762213 , 3.68532, 7.03162, ...$.}
\label{SpikeLPA30}
\end{figure*}

\paragraph*{\tt Tricritical}

The field expansion solution \eqref{solbetas} reveals that in $d_{c}=3$ the couplings $\lambda_{2},\lambda_{4}$
and $\lambda_{6}$ are of order $O(\varepsilon)$ while starting from $\lambda_{8}$ they are all order $O(\varepsilon^{2})$.
Thus the minimal truncation is $n_v=3$, which gives $\alpha=\tfrac{1}{20}$. Inserting this into \eqref{solbetas} gives the following non-zero fixed points values
\begin{equation*}
\lambda_2=\tfrac{\varepsilon}{20}+O(\varepsilon^{2})  \qquad \qquad
\lambda_4=-\tfrac{\varepsilon}{10}+O(\varepsilon^{2}) 
\end{equation*}
\vspace{-0.5cm}
\begin{equation}
\lambda_{6}=\tfrac{\varepsilon}{10}+O(\varepsilon^{2})\,,\label{vapising}
\end{equation}
while the rest are at least of order $O(\varepsilon^2)$.
The stability matrix around the ${\tt Tricritical}$ fixed point is tri-diagonal
\begin{equation*}
\mathbb{M}={\tiny\left(\begin{array}{ccccccc}
-2-\frac{\varepsilon}{10} & -1+\frac{\varepsilon}{20} & 0 & 0 & 0 & 0&\cdots\\
\frac{\varepsilon}{10} & -1-\frac{8\varepsilon}{5} & -1+\frac{\varepsilon}{20} & 0 & 0 & 0&\cdots\\
0 & \frac{3\varepsilon}{2} & -\frac{7\varepsilon}{2} & -1+\frac{\varepsilon}{20} & 0 & 0&\cdots\\
0 & 0 & 7\varepsilon & 1-\frac{29\varepsilon}{5} & -1+\frac{\varepsilon}{20} & 0&\cdots\\
0 & 0 & 0 & 21\varepsilon & 2-\frac{17\varepsilon}{2} & -1+\frac{\varepsilon}{20}&\cdots\\
0 & 0 & 0 & 0 & \frac{99\varepsilon}{2} & 3-\frac{58\varepsilon}{5}&\cdots\\
\vdots & \vdots & \vdots & \vdots & \vdots & \vdots&\ddots
\end{array}\right)}
\end{equation*}
We find the spectrum
\begin{equation}
\theta_2=-2+O(\varepsilon^{2}) \quad \theta_4=-1-\tfrac{\varepsilon}{5}+O(\varepsilon^{2}) \quad \theta_{6}=2\varepsilon+O(\varepsilon^{2}) \nonumber
\end{equation}
\begin{equation} \label{specLPATri}
\theta_8=1+\tfrac{41}{5}\varepsilon+O(\varepsilon^{2}) \qquad \theta_{10}=2+20\varepsilon+O(\varepsilon^{2}) \,, 
\end{equation}
which agrees with perturbation theory \eqref{genspectrum}.
\\

\paragraph*{\tt Multi-critical}

The same procedure works for all other multi-critical theories and completely confirms the general expression \eqref{genspectrum}.

\subsection*{\color{teal}Scaling Solutions}\vspace{-0.2cm}

A signature of all functional RG approaches is the possibility of studying fixed points beyond a field expansion -- i.e., beyond beta functions for a finite set of couplings -- directly at the functional level. This is achieved by setting $\beta_v=0$ in \eqref{flowLPAv}, which yields a nonlinear ODE for the scaling potential $v_*(\varphi)$:
\begin{equation}\label{LPAvss}
0 =-d\,v_*+\tfrac{d-2}{2}\varphi \, v'_*+e^{-v''_*}\,.
\end{equation}
Clearly, this equation admits the trivial Gaussian solution $v_G(\varphi) = \tfrac{1}{d}$ with $v_G'=v_G''=0$ in all dimensions.
%
To find non-trivial scaling solutions, it is convenient to work with the derivative $u(\varphi)\equiv v'_*(\varphi)$, for which the fixed point equation becomes simply
\begin{equation}\label{LPAu}
0=-(d+2) u+ (d-2) \varphi\,  u'-2u'' e^{-u'}\,.
\end{equation}
This can be solved for $u''$ given initial conditions $u(0)=0$ (imposed by $\mathbb{Z}_2$ symmetry) and $u'(0) = \sigma$.
By tuning the parameter $\sigma$, one finds that solutions generally encounter a singularity at some finite $\varphi_s$.
Plotting $\varphi_s$ as a function of $\sigma$ yields a spike plot -- a standard tool in functional RG studies \cite{Morris:1994ki,Morris:1996nx,Hellwig:2015woa} -- which displays the maximum extension of the solution and exhibits sharp spikes at discrete values of $\sigma$, representing functional scaling solutions.
Figure~\ref{SpikeLPA30} shows the Gaussian ($\sigma_G=0$) and the Wilson-Fisher ($\sigma_{\tt Ising} = -0.29754...$) spikes in $d=3$.
These findings fully confirm -- and justify -- the results obtained with the polynomial expansion above.

The $d$-dependence of the scaling solutions reveals the full richness of the FDR-LPA flow across dimensions. Above $d=4$ in integer dimension, we find only the Gaussian solution, in full agreement with the Ginzburg criterion and the expectation that the Gaussian fixed point is infrared-stable for $d>4$. As $d$ is lowered below $d=4$, the Wilson-Fisher spike emerges from the Gaussian one at the critical dimension $d_c=4$, signalling the appearance of the non-trivial Ising universality class. Lowering the dimension further, new multi-critical fixed points appear as additional spikes emanating from the Gaussian solution  at the critical dimensions $d_c=\tfrac{2n}{n-1}$ for $n=2,3,4,...$, corresponding to the Landau-Ginzburg potentials $\varphi^{2n}$ and their associated multi-critical universality classes \cite{Codello:2012sc}. This pattern continues down to $d=2$, where the term proportional to $(d-2)$ in \eqref{LPAu} vanishes and the fixed point equation \eqref{LPAvss} degenerates to $0 =-d\,v_*+e^{-v''_*}$
which can be integrated exactly, yielding the Sine-Gordon model \cite{Nagy:2009pj}.
Taken together, these examples demonstrate that the FDR-LPA flow remains valid and well-defined in all dimensions, clearly overcoming the traditional use of dimensional regularization, which is typically confined to expansions around isolated critical dimensions $d_c$.

\subsubsection*{Eigen-perturbations}\vspace{-0.2cm}

Around a fixed point, the time-dependent dimensionless potential admits the linearized scaling form ($t\equiv\log \mu$)
\begin{equation}
v_t(\varphi) = v_*(\varphi) + e^{\theta t}\,\delta v(\varphi)\,,
\end{equation}
where all the $t$-dependence is carried by the exponential factor and the eigen-perturbation $\delta v(\varphi)$ is a function solely of the field $\varphi$. The critical exponent $\theta$ is the eigenvalue associated with the perturbation, with $\theta<0$ ($\theta>0$) corresponding to a relevant (irrelevant) direction in the RG sense. The eigen-perturbations satisfy the linearized LPA equation, obtained by expanding the flow equation \eqref{flowLPAv} to first order around $v_*$:
\begin{equation}
-d\,\delta v + \tfrac{d-2}{2}\,\varphi\,\delta v' - e^{-v''_*} \delta v'' = \theta\,\delta v\,,
\end{equation}
subject to the normalization condition $\delta v(0)=1$ and the $\mathbb{Z}_2$-symmetry condition $\delta v'(0)=0$. Solving for the second derivative yields the linear second-order ODE
\begin{equation}\label{linear}
\delta v'' - \tfrac{d-2}{2}\,\varphi\,e^{v''_*}\,\delta v' + (d+\theta)\,e^{v''_*}\,\delta v = 0\,,
\end{equation}
where the coefficient function $e^{v''_*}=e^{u'_*}$ is known numerically with high precision from the spike plot analysis of the previous paragraph. 
The linearized equation  \eqref{linear}
can be multiplied by the integrating factor 
\(\mu(\varphi)=\exp\bigl(-\frac{d-2}{2}\int_0^\varphi t\,e^{v_*''(t)}\mathrm{d}t\bigr)\) 
and rewritten as a self-adjoint eigenvalue problem
\begin{equation}\label{linear2}
-\frac{1}{w}(\mu\,\delta v')' - d\,\delta v = \theta\,\delta v\,,
\end{equation}
on the Hilbert space \(L^2\bigl(\mathbb{R},\,w(\varphi)\,\mathrm{d}\varphi\bigr)\) 
with weight \(w = e^{v_*''}\mu\). 
The operator is symmetric on functions for which the boundary term 
\([\mu(\delta v'\,\overline{\delta v} - \delta v\,\overline{\delta v}')]_{-\infty}^{\infty}\) vanishes;
because the weight typically decays like a Gaussian at infinity,
only those solutions that do not blow up faster than a polynomial (in fact, that are square-integrable with respect to \(w\)) 
are admissible. This normalisability condition selects a discrete set of \(\theta\): 
the spectrum is the pure point spectrum of a compact resolvent operator, 
hence quantized. 
The numerical solution to this eigenvalue problem is shown in Figure~\ref{SpikeLPA30}, and the resulting eigenvalues are in perfect agreement with those obtained from the polynomial expansion \eqref{thetai}. This agreement between this approach and the one based on the field expansion constitutes a non-trivial consistency check of the fixed point structure and confirms the reliability of the truncations employed.

\section*{Derivative Expansion $O(\partial^2)$}\label{de2section}\vspace{-0.2cm}

In this section we consider one of the simplest approximations to functional flows capable of estimating the anomalous dimension: the derivative expansion at order $\partial^2$ (FDR-DE2). After deriving the beta functionals for the potential $V$ and the wave-function renormalization $Z$, we use a field expansion to obtain an estimate for the anomalous dimension of the {\tt Ising} universality class, under various approximations. We then discuss how to recover the $\varepsilon$-expansion and show that FDR-DE2 is in fact $\varepsilon$-correct for the {\tt Ising} class below $d_c=4$, a property not shared by most derivative expansions \cite{ODwyer:2007brp,DEvsEE}.

\subsection*{\color{teal}Derivation}\vspace{-0.2cm}

We write the DE2 action as follows
\begin{equation}\label{SDE2}
S[\phi]=\int {\rm d}^d x \biggl\{\frac{1}{2}(\partial\phi)^2 +V(\phi)+Z(\phi)\,\frac{1}{2}(\partial\phi)^2 \biggl\}
\end{equation}
keeping the composite operator $Z(\phi) \frac{1}{2}(\partial\phi)^2$ as a ``perturbation''. 
As usual for any derivative expansion calculation, we will extract the flow of $\beta_Z$ from the two point function. 
More precisely the divergent term of the second functional derivative of \eqref{DEdiv} in momentum space is of the form
\begin{equation}\label{Gamma2div}
\Gamma^{(2)}_{1}(p^2)|_{\infty}=-\frac{1}{\varepsilon}\left(\beta_V''+p^2\beta_Z \right)\,.
\end{equation}
where $d=d_c-\varepsilon$. 
At one-loop $\Gamma_1^{(2)}(p^2)$ will be computed from the Feynman diagrams shown in Figure~\ref{FIGtwopointoneloop} --  the polarization and the tadpole -- but before we need to find the Feynman rules appropriate to our action \eqref{SDE2}.
It is not difficult to show that these are:
\begin{equation}\label{S2}
S^{(2)}(p_1,p_2) = -(1+ Z)(p_1 \cdot p_2)+V''
\end{equation}
\begin{equation}\label{S3}
S^{(3)}(p_1,p_2,p_3)=-Z'(p_2 \cdot p_3 + p_1 \cdot p_2 + p_1 \cdot p_3) +V'''
\end{equation}
\begin{figure}[t]
    \centering
    \begin{tikzpicture}
    \draw[thick] 
    (0,0) circle (\radius*0.5);
       \draw[thick] 
    (0,0) circle (\radius*0.5);
    \draw[thick] 
    (-2*\radius*0.65,0) -- (-\radius*0.5,0);
        
    \draw[thick] 
        (\radius*0.5,0) -- (2*\radius*0.65,0);
       
\begin{scope}[xshift=-120,yshift=10]
      \draw[thick] 
    (0,0) circle (\radius*0.5);
        \draw[thick] 
        (-1,-2*0.5)--(0,-0.5*\radius) ;
     
        \draw[thick] 
        (1,-2*0.5)--(0,-\radius*0.5) ;
        \end{scope}
\end{tikzpicture}
\caption{Feynman diagrams for $\beta_Z$ at one-loop. From left: tadpole ($\frac{1}{2}$) and polarization ($-\frac{1}{2}$).
\label{FIGtwopointoneloop}}
\end{figure}
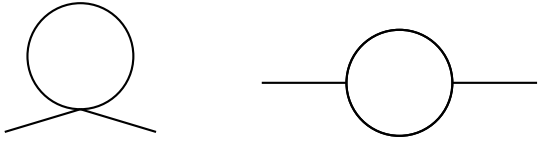
\begin{widetext}
\begin{equation}\label{S4}
S^{(4)}(p_1,p_2,p_3,p_4)=-Z''(p_3\cdot p_4+ p_2 \cdot p_4+p_2 \cdot p_3+p_1 \cdot p_4+p_1 \cdot p_3+p_1  \cdot p_2)+V''''\,,
\end{equation}
\end{widetext}
where now $\phi \to \varphi$ is taken constant.
For the diagrams of Figure~\ref{FIGtwopointoneloop} we will just need the following simpler expressions
\begin{equation}\label{S3bis}
S^{(3)}(q,p,-q-p)=Z'(q^2+q\cdot p+p^2)+V'''
\end{equation}
and
\begin{equation}\label{S4bis}
S^{(4)}(q,p,-q-p)= Z''(p^2+q^2)+V''''\,.
\end{equation}
Finally, the momentum space propagator is given by the inverse of $S^{(2)}(q,-q)$:
\begin{equation}\label{propagador}
G(q^2)=\frac{1}{q^2(1+Z) +V''}\,.
\end{equation}
We first re-compute $\beta_V$ in presence of $Z$ and then move to determine $\beta_Z$.

\subsubsection*{$\beta_V$ in presence of $Z$}\vspace{-0.2cm}

The presence of the wave-function renormalization $Z$ changes $\beta_V$ from its LPA form. 
We need to re-compute the one-loop effective action \eqref{oneloop},
but using \eqref{S2} as input Hessian $S^{(2)}(q,-q)=q^2 (1+Z)+V''$.
Then the calculation proceed as in the LPA case, only that \eqref{steps} is replaced by 
\begin{equation}\label{step1}
\Gamma_{1} = - \tfrac{1}{2}\Omega\int_0^{\infty} \frac{{\rm d}s}{s}\int_q   e^{-s[q^2 (1+Z)+V'']} \,,
\end{equation}
where we expressed the functional trace explicitly as a momentum integral ${\rm Tr} = \int_q \equiv \int\frac{{\rm d}^d q}{(2\pi)^d}$. 
If we rescale the proper-time parameter $s\to s/(1+Z)$ then \eqref{Gamma1} gets transformed to
\begin{equation}
\Gamma_{1}= -\Omega\frac{\Gamma\big(-\frac{d}{2}\big)}{2(4\pi)^{\frac{d}{2}}} 
\left(\frac{V''}{1+Z}\right)^{\frac{d}{2}}\,.\vspace{0.1cm}
\end{equation}
As before, the Euler Gamma function presents poles when $d=2n\equiv d_c$ with integer $n$ with residues given in \eqref{polesGamma}. 
The DR beta functionals that generalize \eqref{betaVDR} are
\begin{equation}
\beta^{\rm DR}_V(d_c) =  \frac{(-1)^{\frac{d_c}{2}}}{(4\pi)^\frac{d_c}{2}\big(\frac{d_c}{2}\big)!}
\left(\frac{V''}{1+Z}\right)^{\frac{d_c}{2}} \,.
\end{equation}
\\
Finally, we apply the master formula \eqref{master} and sum over all critical dimensions $d_c=0,2,4,6,8,...$
to obtain the following beta functional for the potential in presence of the wave-function renormalization 
\begin{equation}\label{betaVZ}
\beta_V(d)=\mu^{d}\,e^{-\tfrac{V''}{4 \pi  \mu^2 (1+Z)}}\,.
\end{equation}
To conclude, the dimensionless flow \eqref{dimlessbetav} that follows is
\begin{equation} \label{betavz0}
\beta_v = -d\, v+\tfrac{d-2+\eta}{2}\varphi \,v'+e^{-\tfrac{v''}{1+z}} \,,
\end{equation}
where we have rescaled away the usual $4\pi$'s.
Note that the only correction with respect to the LPA flow \eqref{flowLPAv} is the presence of the denominator $1+z$ in the exponential.

\subsubsection*{Computing $\beta_Z$}\vspace{-0.2cm}

Diagrammatically the second functional derivative of the one-loop effective action has two contributions as shown in Figure~\ref{FIGtwopointoneloop}.
The first one (the tadpole) will have a contribution of the form
\begin{eqnarray}\label{tadpole}
{\rm tadpole} &=& \tfrac{1}{2}\int_q S^{(4)}(p,-p,q,-q)\, G(q^2) \nonumber \\ 
&=& \tfrac{1}{2} \int_q \left[ (q^2 +p^2)Z''+ V''''  \right]G(q^2)\,,
\end{eqnarray}
where we used \eqref{S4bis}. The polarization term is
\begin{equation*}
{\rm polarization} = -\tfrac{1}{2}\int_q \big[S^{(3)}(q,p,-q-p) \big]^2 G\big((q+p)^2\big)G(q^2)\,,
\end{equation*}
which using \eqref{S3bis} and expanding becomes
\begin{widetext}
\begin{eqnarray}\label{polarization}
 {\rm polarization} &=& -\tfrac{1}{2}\int_q \big[(V''')^2+2Z'V'''(q^2+q\cdot p+p^2)
+(Z')^2(q^2+q\cdot p+p^2)^2\big]G\big((q+p)^2\big)G(q^2)\,.
\end{eqnarray}
\end{widetext}
From \eqref{tadpole} and \eqref{polarization} we can identify four sources of $p^2$ contributions.
Of the following, the first comes from the tadpole, while the other three from the polarization:
\\
\vspace{-0.5cm}
%
\begin{eqnarray}\label{integrals}
(a)&& \frac{1}{2}Z''p^2 \int_q G(q^2) \\
(b)&& -\frac{1}{2} (V''')^2 \int_q G\big((q+p)^2\big)G(q^2)\nonumber\\
(c)&& -Z'V''' \int_q \Big[ q^2+ p\cdot q+p^2 \Big] G\big((q+p)^2\big)G(q^2)\nonumber\\
(d)&& -\frac{1}{2} (Z')^2 \int_q \Big[q^4+(p\cdot q)^2+2q^2(p\cdot q)+2p^2q^2 \nonumber\\&&\qquad\qquad\qquad+ 2(p\cdot q) p^2 + p^4\Big] G\big((q+p)^2\big)G(q^2)\,.\nonumber
\end{eqnarray}
%
As usual, the momentum integral $q$ splits into an angular and a radial integral,
\begin{equation*}
\int_q \rightarrow \frac{S_{d-1}}{(2\pi)^d} \int_{-1}^1 {\rm d}x\, (1-x^2)^{\frac{d-3}{2}}  \int_0^\infty {\rm d}q\, q^{d-1}\,,
\end{equation*}
where $x=\cos \theta$, $\theta$ being the angle between $p$ and $q$.
We recall that $S_d=\tfrac{2\pi^{d/2}}{\Gamma(d/2)}$ is the spherical area.
Furthermore, using $(q+p)^{2} = q^{2}+p^{2}+2 p q x$, we expand the $p$-dependence of the propagators as
\begin{eqnarray}
G\big({\tiny(q+p)^{2}}\big)	
&=&	G(q^{2})+2qx\,G'(q^{2})\,p \label{expprog}\\
&&+\big[G'(q^{2})+2q^{2}x^{2}\,G''(q^{2})\big]\,p^{2}+O(p^{3})\nonumber\,.
\end{eqnarray}
In the complex $d$-plane the integrals \eqref{integrals} are all well defined, and after inserting \eqref{expprog} we can chase, case by case, their total $p^2$ contributions.
After the $x$-integration, all integrals left over are Gaussian -- with powers of the internal momenta -- and can be performed straightforwardly. The final results are:
\begin{eqnarray}\label{contributions}
(a) &=& \frac{(V'')^{\frac{d}{2}-1}}{(4\pi)^{\frac{d}{2}} (1+Z)^{\frac{d}{2}}  \Gamma(\frac{d}{2})} \tfrac{\pi}{2}  \csc \tfrac{\pi d}{2}\nonumber\\
(b) &=& \frac{ (d-4) (d-2)   (V'')^{\frac{d}{2}-3}   }{24 (4\pi)^{\frac{d}{2}}(1+Z)^{\frac{d}{2}-1} \Gamma(\frac{d}{2})}  \tfrac{\pi}{2}\csc\tfrac{\pi d}{2}\nonumber\\
(c) &=& -\frac{(d-10) (d-2)  (V'')^{\frac{d}{2}-2} }{12(4\pi)^{d/2} (1+Z)^{\frac{d}{2}} \Gamma(\frac{d}{2})}  \tfrac{\pi}{2}  \csc \tfrac{\pi d}{2}\nonumber\\
(d) &=& \frac{(d^2-18 d-4)(V'')^{\frac{d}{2}-1}}{24 (4\pi)^{\frac{d}{2}}(1+Z)^{\frac{d}{2}+1} \Gamma (\frac{d}{2})}\tfrac{\pi}{2}  \csc \tfrac{\pi  d}{2}\,.
\end{eqnarray}
These expressions all have divergences caused by the same trigonometric factor
\vspace{0.2cm}
\begin{equation}
\tfrac{\pi}{2} \csc\tfrac{\pi d}{2} =
(-1)^{\frac{d_c}{2}+1}\big\{\tfrac{1}{\varepsilon }+\tfrac{\pi ^2}{24}\varepsilon+O(\varepsilon^2)\big\}\,,
\end{equation}
\\
where $d_c$ runs over all even positive integers (including zero), as in the case of the potential.
For each term in \eqref{contributions} we first extract the $\tfrac{1}{\varepsilon}$-pole and then, using \eqref{Gamma2div}, we deduce their contribution to the DR beta functional of the wave-function renormalization.
Putting all together we finally find the following general expression:
\begin{widetext}
\begin{eqnarray}\label{BZDR}
\beta_Z^{\rm DR}(d_c) &=& \frac{(-1)^{\frac{d_{c}}{2}}}{(4\pi)^{\frac{d_{c}}{2}}\Gamma(\frac{d_{c}}{2})}
\left\{  \frac{ (d_{c}-4) (d_{c}-2)  (V'')^{\frac{d_{c}}{2}-3} }{24 (1+Z)^{\frac{d_{c}}{2}-1}} (V''')^2
+\frac{(V'')^{\frac{d_{c}}{2}-1}}{ (1+Z)^{\frac{d_{c}}{2}}} Z'' 
\right. \nonumber\\
&&\left. \qquad\qquad\qquad\quad
-\frac{(d_{c}-10) (d_{c}-2)  (V'')^{\frac{d_{c}}{2}-2} }{12(1+Z)^{\frac{d_{c}}{2}} }Z'V'''
+\frac{(d_{c}^2-18 d_{c}-4)(V'')^{\frac{d_{c}}{2}-1}}{24
(1+Z)^{\frac{d_{c}}{2}+1}}(Z')^2 
\right\}\,.
\end{eqnarray}
\end{widetext}
One can make some checks at this point by evaluating this expression in $d_c=2$, $d_c=4$ or $d_c=6$ and compare with known perturbative results. In $d_c=4$ equation \eqref{BZDR} becomes
\begin{equation}
\beta^{\rm DR}_Z(4) = \frac{V''' Z'}{(4\pi)^2}+\frac{V'' Z''}{(4\pi)^2}
-\frac{5 V'' (Z')^2}{2(4\pi)^2}\,,
\end{equation}
which agrees with the results of \cite{ODwyer:2007brp,Codello:2017hhh}.
In $d_c=6$ again only $\beta_Z$ gets modified with three terms in addition to the traditional one proportional to the potential
\begin{eqnarray*}
\beta^{\rm DR}_Z(6) &=& -\frac{(V''')^2}{6(4\pi)^3}-\frac{(V'')^2 Z''}{2(4\pi)^3}\\
&&-\frac{2 V'' V''' Z'}{3(4\pi)^3}+\frac{19 (V'')^2 (Z')^2}{12(4\pi)^3}\,.
\end{eqnarray*}
It is instructive to consider also the $d_c=2$ case which has been discussed in \cite{Baldazzi:2020vxk}
\begin{equation}
\beta^{\rm DR}_Z(2) = -\frac{Z''}{4\pi}+\frac{3 (Z')^2}{8\pi}\,.
\end{equation}
Note that in all these expressions for $\beta^{\rm DR}_Z(d_c)$, the power $\tfrac{d_c}{2}$ to which the factors of $4\pi$ are raised reflects the critical dimension.

The summation in the master formula \eqref{masterZ} for $\beta_Z$ is easily done and gives the following beta functional
\begin{widetext}
\begin{eqnarray}\label{betaZ}
\beta_Z &=&
\left\{
-\frac{1}{6}\frac{\mu^{d-6}}{(4\pi)^3 }\frac{(V''')^2}{(1+Z)^2}  -\frac{\mu^{d-2}}{(4\pi)}\frac{Z''}{1+Z}\right.
+\left[\frac{\mu^{d-4}}{(4\pi)^2}+\frac{1}{3}\frac{\mu^{d-6}}{(4\pi)^3}\frac{V''}{1+Z}\right] \frac{V'''Z'}{(1+Z)^2}
\nonumber\\&&\left.
\qquad\qquad\qquad\qquad\qquad
+\left[
\frac{3}{2}\frac{\mu^{d-2}}{(4\pi)}
-\frac{\mu^{d-4}}{(4\pi)^2}\frac{V''}{1+Z}
-\frac{1}{6}\frac{\mu^{d-6}}{(4\pi)^3}\frac{(V'')^2}{(1+Z)^2}
\right]\frac{(Z')^2}{(1+Z)^2}
\right\} e^{ -\tfrac{V''}{4 \pi \mu^2 (1+Z)}} \,.
\end{eqnarray}
\end{widetext}
Clearly the $(V''')^2$ term reproduces the LPA expression given in \cite{Beretta:2026zcy} in the limit $Z\to 0$
\begin{equation}\label{betaZLPA}
\beta_Z^{\rm LPA} =  -\frac{\mu^{d-6}}{6(4\pi)^3}(V''')^2 e^{ -\frac{V''}{4 \pi \mu^2}} 
\end{equation}
Equations \eqref{betaVZ} and \eqref{betaZ} constitute the FDR-DE2 flow and are the main result of this paper.

To gauge the influence of various aspects of our formalism, we now discuss three approximations of the DE2 beta functionals \eqref{betaZ}, which we term the \textit{strict}, \textit{light}, and \textit{light-strict} forms.
The \textit{strict} approximation, as formulated in \cite{DePolsi:2020pjk}, consists of discarding terms of fourth order in momenta (either external $p$ or internal $q$ or combinations) and higher when working at order DE2. The discarded terms can be traced to the polarization diagram, and in particular to contribution $(d)$ of \eqref{integrals}. 
The {\it strict} prescription therefore amounts to setting terms proportional to $(Z')^2$ to zero. This step must be performed before expanding the propagator and applies only to products of vertices.
The \textit{light} approximation consists of setting $Z=0$ inside the propagator \eqref{propagador}, thereby alleviating the algebraic complexity of the beta functional. The \textit{light-strict} case is simply the combination of the two: one simultaneously drops the $(Z')^2$ terms (strict) and sets $Z=0$ in the propagator (light). These are the ``perturbative'' equations studied in \cite{Beretta:2026zcy} in which we retained only linear terms in $Z$.  

Finally, we report the dimension-less form of \eqref{betaZ}, in which we also rescale away the $4\pi$'s via a field and function redefinition:
\begin{widetext}
\begin{eqnarray}
\beta_z \!&=&\!  \eta\, (1+z)\!+\!\tfrac{d-2+\eta}{2}\varphi z'\!- \!\left\{\frac{(v''')^2}{6 (1+z)^2} 
\!+\!\frac{z''}{1+z} \!- \! \left[1\!+\!\frac{v''}{3(1+z)}  \right] \! \frac{v''' z'}{(1+z)^2} \right.  
\left.\!-\!\left[\frac{3}{2}\!-\!\frac{v''}{1+z} \!-\!\frac{(v'')^2}{6(1+z)^2}\right] \!\frac{(z')^2}{(1+z)^2}
\right\} e^{-\frac{v''}{1+z}}\,. \label{betavz} \nonumber\\&&
\end{eqnarray}
\end{widetext}
In the following, and in future works, we will show that a multitude of results in critical phenomena in diverse dimensions can be extracted from this equation together with \eqref{betavz0}.
%
\begin{figure*}
\centering
\includegraphics[width=0.95\columnwidth]{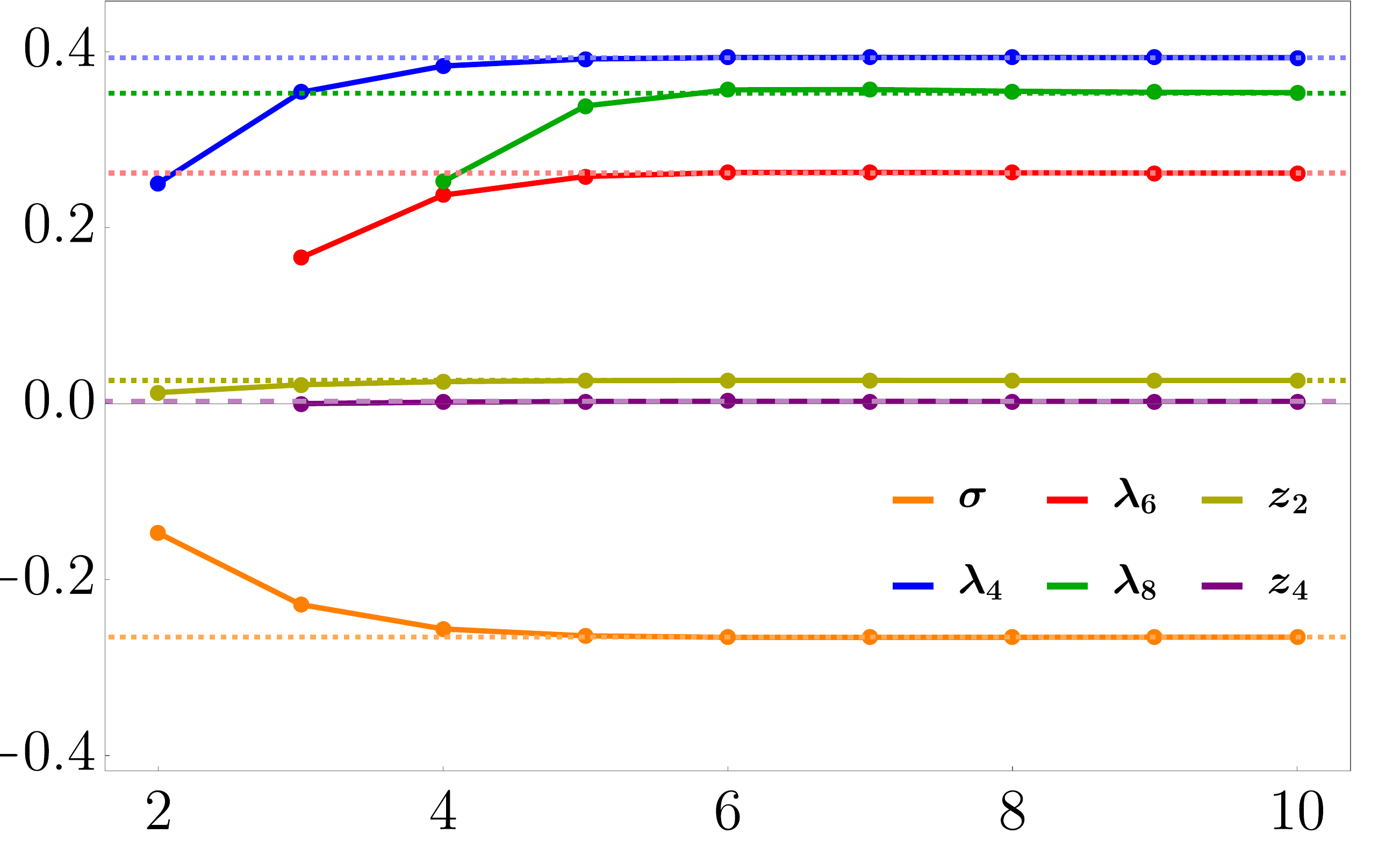}\qquad\qquad\includegraphics[width=0.95\columnwidth]{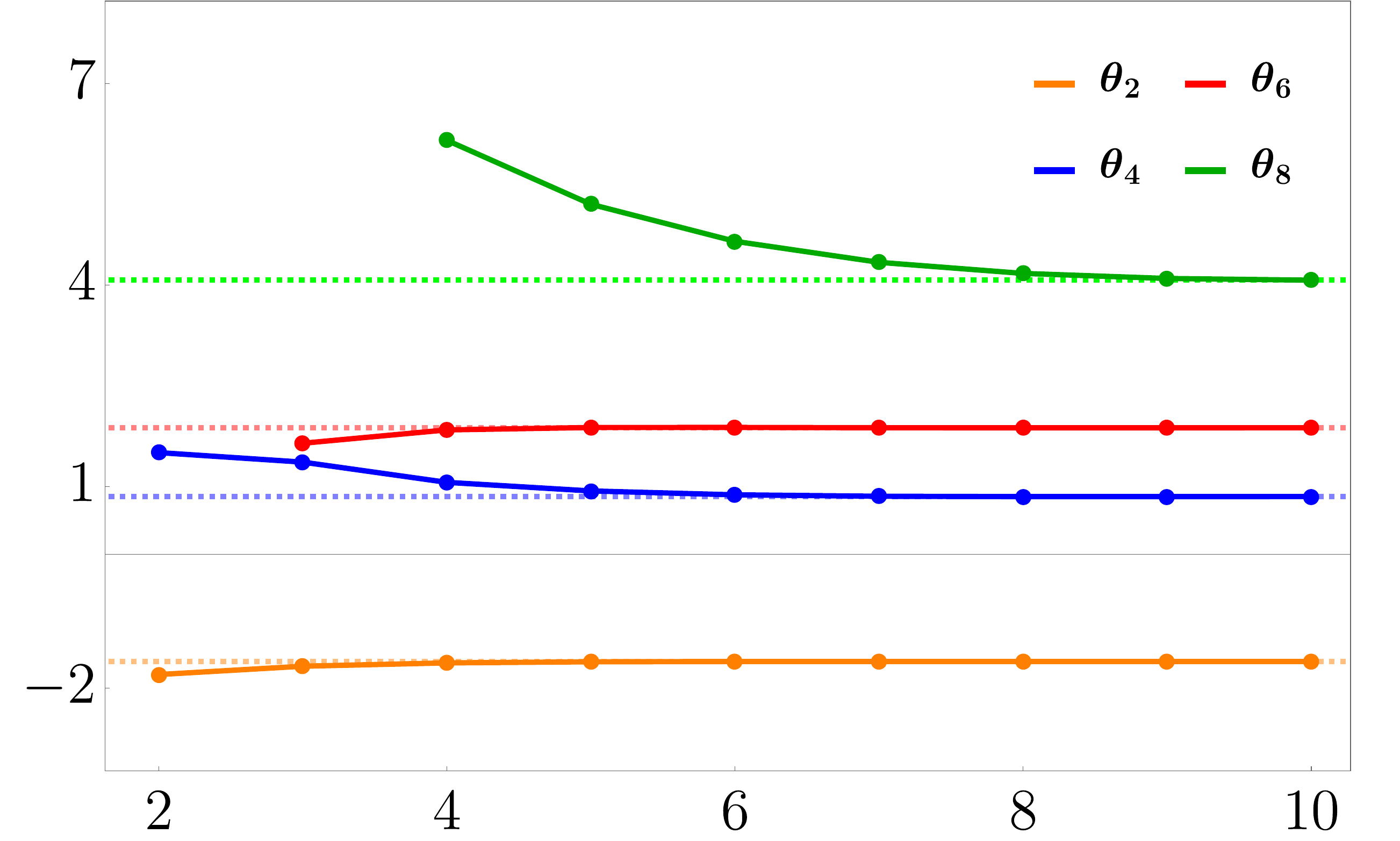}
\caption{Left: Wilson-Fisher fixed point values for the first couplings of the field expansion in $d=3$. Convergence is very rapid, in particular we find $\sigma_{\tt Ising} = -0.2654...$ and $\eta_{\tt Ising} = 0.03425...$. Right: FDR-DE2 spectrum in $d=3$ obtained from the field expansion and relative stability matrix.\label{FIG_sigma_lambda_zeta}}
\end{figure*}

\subsection*{\color{teal}Field Expansion (around $\varphi=0$)}\vspace{-0.2cm}

We start analyzing the DE2 flow by a field expansion around the origin. 
We complement the expansion of the Taylor potential \eqref{taylor} with the following one for the wave-function renormalization
\begin{equation}
z(\varphi) = \sum_{n=1}^{n_v-1} \frac{z_{2n}}{(2n)!}  \varphi^{2n} = \tfrac{z_2 }{2} \varphi ^2+\tfrac{z_4 }{4!}\varphi ^4   + ... \,.\label{taylorz}
\end{equation}
Note that we have chosen -- quite arbitrarily -- to truncate this series at $n_v-1$ where $n_v$ is the truncation order of the potential.
Inserting now the Taylor expansions in the respective beta functional -- \eqref{taylor} in \eqref{betavz0} and \eqref{taylorz} in \eqref{betavz} -- then expanding and matching coefficients, gives the following coupled system of beta functions representing the FDR-DE2 flow:
\begin{widetext}
\begin{eqnarray}
\beta_{2} & =&(-2+\eta)\lambda_{2}+(\lambda_{2}z_{2}-\lambda_{4})e^{-\lambda_{2}}\nonumber\\
\beta_{4} & =&(d-4+2\eta)\lambda_{4}+3\left(\lambda_{4}-\lambda_{2}z_{2}\right)^{2}e^{-\lambda_{2}}+(-\lambda_{6}-6\lambda_{2}z_{2}^{2}+6\lambda_{4}z_{2}+\lambda_{2}z_{4})e^{-\lambda_{2}}\nonumber\\
\beta_{6} & =&(2d-6+3\eta)\lambda_{6}+15\left(\lambda_{2}z_{2}-\lambda_{4}\right)^{3}e^{-\lambda_{2}}-15\left(\lambda_{2}z_{2}-\lambda_{4}\right)\left(\lambda_{6}+6\lambda_{2}z_{2}^{2}-6\lambda_{4}z_{2}-\lambda_{2}z_{4}\right)e^{-\lambda_{2}}\nonumber\\
&&+ \left(-\lambda _8+90 \lambda _2 z_2^3-90 \lambda _4 z_2^2+15 z_2 \left(\lambda _6-2 \lambda _2 z_4\right)+\lambda _2 z_6+15 \lambda _4 z_4\right)e^{-\lambda _2}\nonumber\\
\beta_{8} & =&... \label{betaszv-v}
\end{eqnarray}
and
\begin{eqnarray}
\beta_{z_{2}} & =&(d+2\eta-2)z_{2}+\tfrac{1}{3}  \left(\left(2 \lambda _2+9\right) \lambda _4 z_2-\left(\lambda _2^2+9 \lambda _2-12\right) z_2^2-\lambda _4^2-3z_4\right)e^{-\lambda _2}\nonumber\\
\beta_{z_{4}} & =&(2d+3\eta-4)z_{4}-2\left(\lambda_{2}z_{2}-\lambda_{4}\right)^{3}e^{-\lambda_{2}}
+\tfrac{1}{3}\left(4\lambda_{6}-21\lambda_{2}z_{2}^{2}+21\lambda_{4}z_{2}-4\lambda_{2}z_{4}\right)\left(\lambda_{2}z_{2}-\lambda_{4}\right)e^{-\lambda_{2}}\nonumber\\
 & &+\left(66\lambda_{2}z_{2}^{3}-66\lambda_{4}z_{2}^{2}+5z_{2}\left(\lambda_{6}-3\lambda_{2}z_{4}\right)+10\lambda_{4}z_{4}\right)e^{-\lambda_{2}}+\left(19z_{2}z_{4}-42z_{2}^{3}-z_{6}\right)e^{-\lambda_{2}}\nonumber\\
\beta_{z_{6}} & =...  \label{betaszv-z} 
\end{eqnarray}
\end{widetext}
From the condition $z(0)=0$ when $\beta_z =0$ we find the following simple expression for the anomalous dimension\footnote{We have checked that the results do not depend on $\alpha$ if we replace the condition $z(0)=0$ with $z(0)=\alpha$. This follows from the constant reparametrization invariance 
($\varphi \to \alpha \varphi$  and $Z \to Z /\alpha^{2}$) of our FDR-DE2 equations  \cite{Comellas:1997ep,Osborn:2009vs}.}
\begin{equation}\label{etaVZ}
\eta=z_{2}\,e^{-\lambda_{2}}\,.
\end{equation}
As in the LPA case we can solve the system of beta functions iteratively in terms now of just $\sigma$ and $\eta$.
In fact, also the couplings of the wave-function  renormalization are all linear in a coupling $\beta_{z_{2n}}\sim z_{2n+2}$.
Thus we can reduce all to $\sigma=\lambda_2$ and $z_2$,
and then use \eqref{etaVZ} to trade this last coupling for the anomalous dimension.
The general solution has the following form:
\begin{eqnarray}
\lambda_{2}^{*} & {\color{black}=}&\sigma\nonumber\\
\lambda_{4}^{*} & {\color{black}=}&2(\eta-1)e^{\sigma}\sigma\nonumber\\
\lambda_{6}^{*} & {\color{black}=} &-\tfrac{1}{3}\sigma e^{2 \sigma}\big\{6{\color{teal}(d-4)}-48\eta^2+\eta(78-9d)
\nonumber\\&&+\sigma^2(\eta-2)^2-18\sigma(\eta-1)(\eta-2)\big\}\nonumber\\
\lambda_{8}^{*} & =&...
\label{solZVlambda}
\end{eqnarray}
and
\begin{eqnarray}
z_{2}^{*} & {\color{black}=}&\eta e^{\sigma}\nonumber\\
z_{4}^{*} & {\color{black}=}&-\tfrac{1}{3} e^{2 \sigma } \big\{\eta  \left(-3 (d-2)-4 \sigma ^2+18 \sigma \right)\nonumber
\\&&+\eta ^2 \left(+\sigma ^2-9 \sigma -18\right)+4 \sigma ^2\big\}\nonumber\\
z_{6}^{*} & =&...
\label{solZVzeta}
\end{eqnarray}
The space of solutions is thus two dimensional and this makes the functional analysis of scaling solution more complex.
For this reason we will leave it to the future work \cite{FDR2d} and  focus here on the results we can obtain with the field expansion.

\subsubsection*{Wilson-Fisher fixed point in $d=3$ (II)}\vspace{-0.2cm}

The aim of this section is to study the Wilson-Fisher fixed point in $d=3$ using the DE2 system of beta functions  \eqref{betaszv-v} and \eqref{betaszv-z}.
The procedure follows closely the one explained in {\it Wilson-Fisher fixed point in $d=3$ (I)},
with the main difference that now we can evaluate the anomalous dimension using \eqref{etaVZ} directly from the fixed point values.

Just to explain the procedure we review the first non-trivial truncation $n_v=2$. 
We have three couplings, $\lambda_2$, $\lambda_4$ and $z_2$, whose values at the fixed point are
\begin{equation*}
\lambda^{*}_2=-0.14721 \qquad \lambda^{*}_4=0.2504 \qquad z^*_2=0.0125\,.
\end{equation*}
The anomalous dimension is obtained by evaluating \eqref{etaVZ} at the fixed point $\eta=0.01450$.
The stability matrix evaluated at this fixed point is
\begin{equation*}
\mathbb{M}= \left(
\begin{array}{ccc}
 -1.67656 & -0.273483 & 0.0144108 \\
 -1.15861 & 0.869804 & -0.151361 \\
 -0.341132 & 2.60512 & 2.02863 \\
\end{array}
\right)\,,
\end{equation*}
whose eigenvalues are $\{-1.79523,\, 1.50855\pm 0.35799\,i\}$.
Thus the critical exponents of interest are then $\nu_{n_v=2}=0.5570$ and $\omega_{n_v=2}={\rm Re}\,\theta=1.50855$.
The procedure continues by increasing the truncation number $n_v$. 
We find rapid convergence, $n_v=10$ is already sufficient to fix the first couplings and the low-lying RG eigenvalues, which are shown in Figure~\ref{FIG_sigma_lambda_zeta}.
In particular, we find the final values $\sigma_{\tt Ising} = -0.2654...$ and $\eta_{\tt Ising}  = 0.03425...$, from which we can deduce the critical coordinates of all other couplings using  \eqref{solZVlambda} and \eqref{solZVzeta}.
The final estimates for the critical exponents are $\nu_{n_v=12} = 0.624376$ and $\omega_{n_v=12}=0.855438$.
Finally, to gauge the contributions from the various terms appearing in the r.h.s. of $\beta_Z$ and to judge the effect of the presence of $Z$ in the propagator, we repeat the computations under approximations described in the previous section:  {\it strict}, {\it light}, and {\it light-strict}. 

\begin{figure}
\centering
\includegraphics[width=0.95\columnwidth]{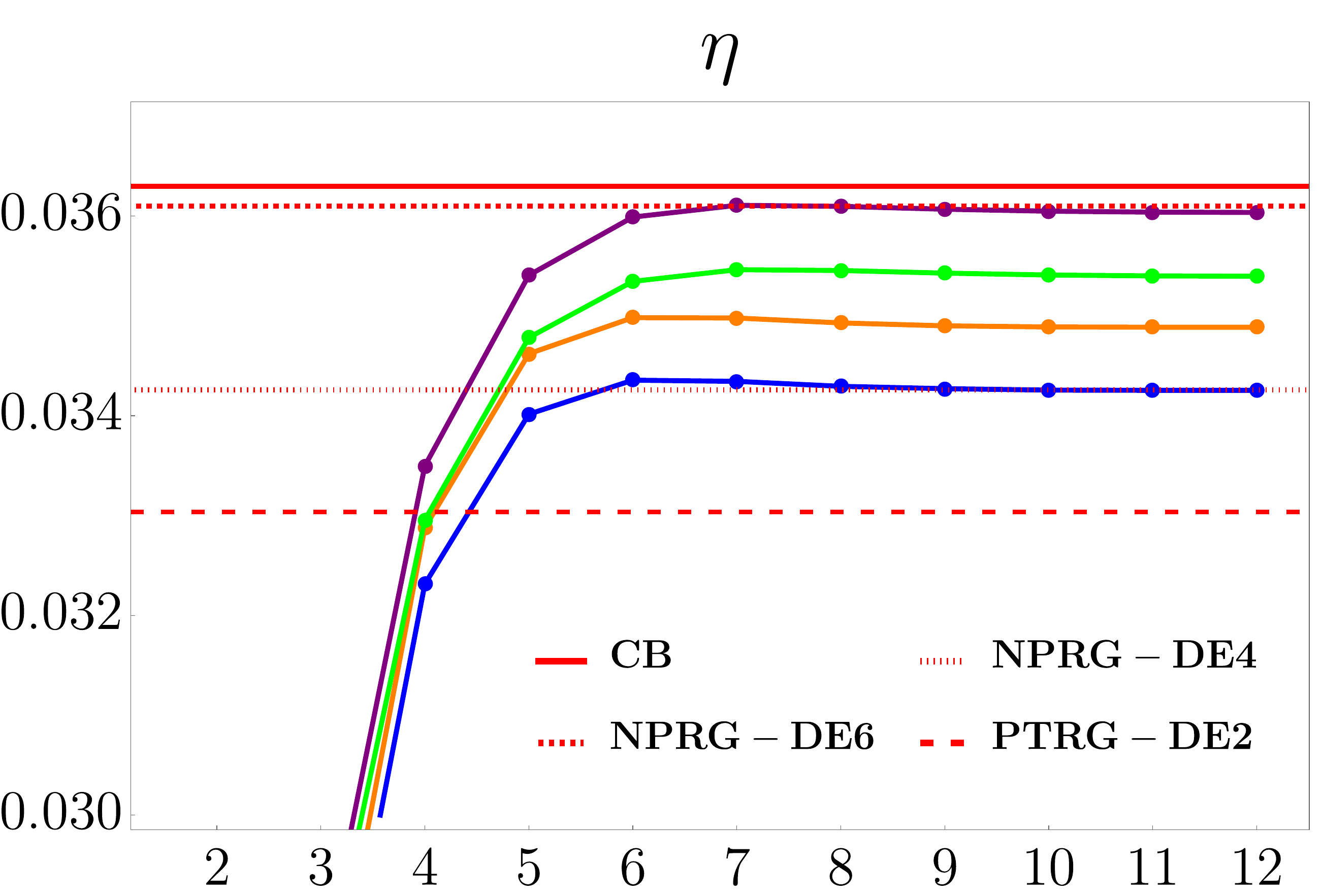}
\caption{Convergence of the critical exponent $\eta$ in the FDR-DE2 approximation (blue/full, orange/strict, green/light, purple/light-strict) in $d=3$. Comparison is made with the state-of-the-art Conformal Bootstrap value in red, NPRG-DE4 (raw-$\Theta^8$) in thin red-dotted, NPRG-DE6 (raw-$\Theta^8$) in thick red-dotted and PTRG-DE2 in red-dashed (computed from equation \eqref{PTRGbetaZ}).}
\label{EtaFDRDE2}
\end{figure}

The convergence of the critical exponent $\eta$ in three dimensions is illustrated in Figure~\ref{EtaFDRDE2}, which reports the FDR-DE2 estimates for the four implementation variants. Among these, the {\it light-strict} (purple) approximation closely tracks the NPRG-DE6 (thick red-dotted) trajectory and converges nearest to the Conformal Bootstrap benchmark (solid red), whereas the {\it full} (blue) variant displays a behavior analogous to the NPRG-DE4 (thin red-dotted) level. The {\it strict} (orange) and {\it light} (green) implementations fall in between these two bounds, thereby forming a consistent hierarchy. As a result, the entire set of FDR-DE2 truncations is effectively sandwiched between the predictions of the fourth- and sixth-order NPRG derivative expansions, indicating that the present second-order scheme encapsulates physics spanning across these higher-order NPRG levels, irrespective of the specific approximation choice. By contrast, the PTRG-DE2 estimate (red-dashed), computed from equation \eqref{PTRGbetaZ}, yields a somewhat weaker prediction, lying further from the Conformal Bootstrap reference. Taken together, these comparisons suggest that the FDR-DE2 framework offers a robust and competitive description of the Wilson-Fisher fixed point in $d=3$, with the modest spread among its variants providing a useful empirical indicator of the residual systematic uncertainty.

\begin{figure*}
\centering
\includegraphics[width=0.95\columnwidth]{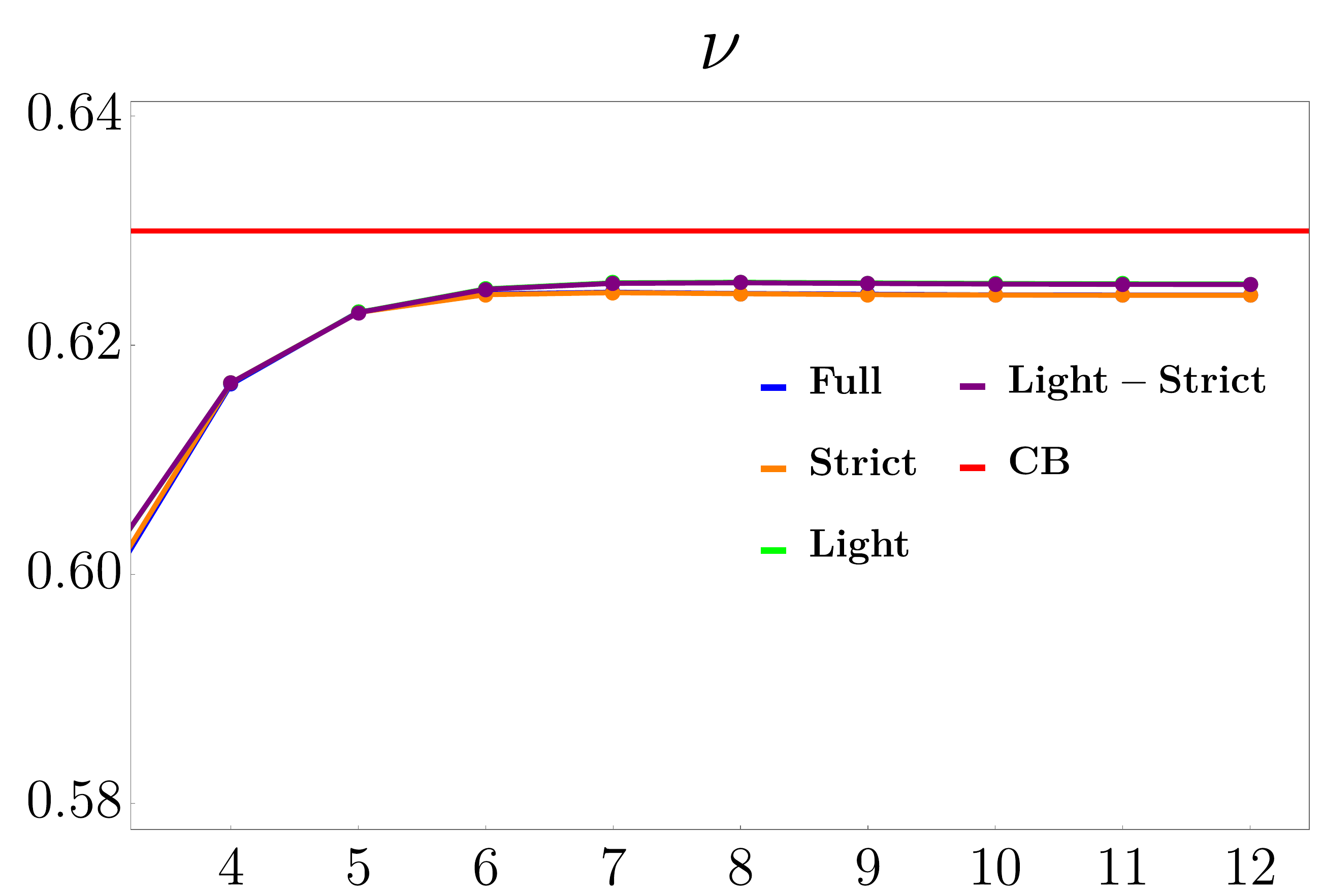}\qquad\qquad
\includegraphics[width=0.95\columnwidth]{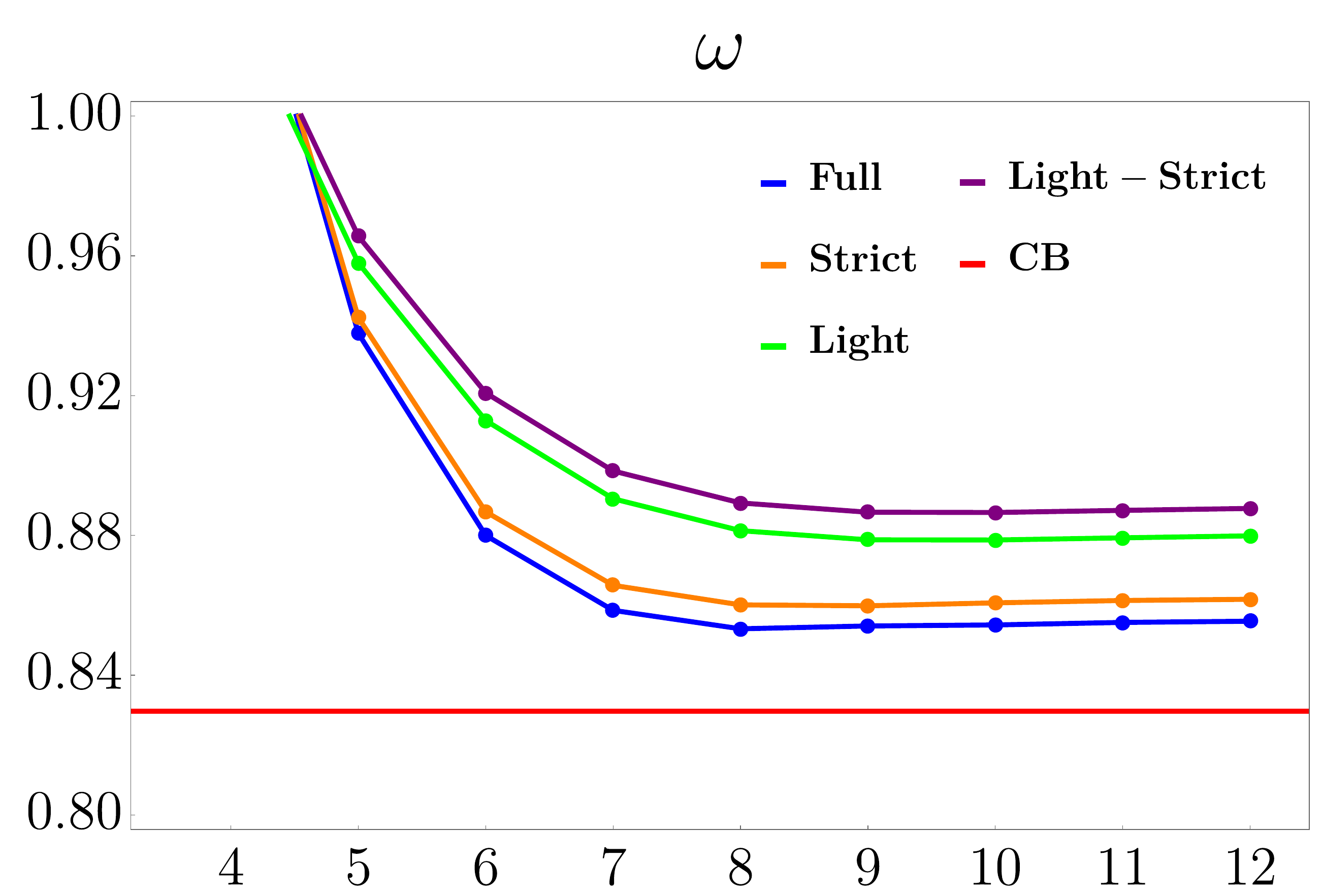}
\caption{Convergence of the critical exponent $\nu$ and $\omega$ in the FDR-DE2 approximation in $d=3$. The state-of-the-art Bootstrap value is shown in red.}
\label{NuOmegaFDRDE2}
\end{figure*}

Figure~\ref{NuOmegaFDRDE2} reports the results for the other exponents.
For the correlation-length exponent $\nu$, the spread between the four FDR-DE2 implementations remains remarkably small, indicating that probably a different source of correction is missing. This is corroborated by the slight deterioration in accuracy observed when compared to the simpler LPA level, which suggests that the inclusion of derivative couplings does not systematically improve all scaling dimensions at this truncation order. In contrast, the situation is reversed for the correction-to-scaling exponent $\omega$: here, the full (blue) implementation emerges as the most accurate, while the light-strict (purple) variant yields the least favorable estimate.

\subsubsection*{Recovering the $\varepsilon$-expansion (II)}\vspace{-0.2cm}

We now perform a general analysis of the $\varepsilon$-expansion, as we did before in the LPA context, for each {\tt multi-critical} universality class with $d_c=\tfrac{2n}{n-1}$ and $n=2,3,4,...$. 
With the inclusion of the wave-function renormalization $Z$, the spectrum splits into two sets -- $\theta^V_i$ and $\theta^Z_i$, the former being the same as in the LPA case. For $\theta^Z_i$ the general formula obtained from perturbation theory \cite{ODwyer:2007brp,Codello:2017hhh} is 
\begin{equation}\label{thetaZ}
-\theta^{Z}_{i}= -i\tfrac{d-2}{2}-2(n-1)\tfrac{n!}{(2n)!}\tfrac{(i+1)!}{(i+1-n)!}\,\varepsilon + O(\varepsilon^2)\,.
\end{equation}
In what follows we will study again the {\tt Ising} and the {\tt Tricritical} universality classes, and then comment the general case.
The main difference is that we have to make ansatzs for both $\sigma$ and $\eta$ -- knowing that the anomalous dimension is of order $O(\varepsilon^2)$. We thus choose the following parametrization: $\sigma = \alpha\, \varepsilon + O(\varepsilon^2)$ and $\eta = \gamma\, \varepsilon^2+ O(\varepsilon^3)$.
\\

\paragraph*{\tt Ising}

Performing an analysis similar to the one done in the LPA case, but now starting from the solution \eqref{solZVlambda} and \eqref {solZVzeta}, and working around $d_{c}=4$, we find $\alpha=-\tfrac{1}{6}$ and $\gamma=\tfrac{1}{54}$.
Amazingly, this gives the correct two-loop result for the anomalous dimension, as already noted in \cite{Beretta:2026zcy}.
The fixed point to linear order is
\begin{equation*}
\lambda_{2}=-\tfrac{\varepsilon}{6}+O(\varepsilon^{2}) \qquad \lambda_{4}=\tfrac{\varepsilon}{3}+O(\varepsilon^{2})
\qquad \lambda_{6}=O(\varepsilon^{3})\,,
\end{equation*}
which is the same we found in the LPA case \eqref{LPAepsilonIsing}. 
For the coupling coming from the $Z$ function, we find the following fixed point
\begin{equation*}
z_{2}=\tfrac{\varepsilon^{2}}{54}+O(\varepsilon^{3})\,,
\end{equation*}
while all other couplings are of order $O(\varepsilon^{3})$.
As noted before, the spectrum is divided in two sectors, one for $V$-direction and one for the $Z$-direction. 
The spectrum regarding $V$ remains unchanged, as expected, with respect to the one computed in the LPA \eqref{specIsinglpa}.
The new thing is the spectrum regarding $Z$ for which we get
\begin{equation*}
\theta_2^{Z}=2 \qquad \theta_4^{Z}=4+\tfrac{4}{3}\varepsilon  \qquad \theta_6^{Z}= 6+4\varepsilon
\end{equation*}
\begin{equation*}
\theta_8^{Z}=8+8\varepsilon \qquad \theta_{10}^{Z} =10+\tfrac{40}{3}\varepsilon\,,
\end{equation*}
and has a two fold degeneracy as noted by \cite{ODwyer:2007brp,Codello:2017hhh}.
It clearly agrees with \eqref{thetaZ}.
\\

\paragraph*{\tt Tricritical}

We again assume $\sigma=\alpha\,\varepsilon$ and $\eta=\gamma\,\varepsilon^{2}$ and
solve around $d_{c}=3$. We find $\alpha=\tfrac{1}{20}$ and $\gamma=\tfrac{1}{450}$. 
In this case we fail to find the right value of $\eta$, since the one predicted by perturbation theory is $\tfrac{\varepsilon^{2}}{500}$ \cite{ODwyer:2007brp,Codello:2017hhh} whereas the one we obtain is $\tfrac{\varepsilon^{2}}{450}$. 
We comment on this in a moment.
The fixed point solution for the potential couplings to order $O(\varepsilon^2)$ is
\begin{equation*}
\lambda_{2}=\tfrac{\varepsilon}{20}+O(\varepsilon^{2}) \qquad\quad \lambda_{4}=-\tfrac{\varepsilon}{10}+O(\varepsilon^{2})
\end{equation*}
\begin{equation*}
\lambda_{6}=\tfrac{\varepsilon}{10}+O(\varepsilon^{2})\qquad\quad \lambda_{8}=-\tfrac{7}{20}\varepsilon^2+O(\varepsilon^{3})\,,
\end{equation*}
with all other couplings of order at least $O(\varepsilon^{2})$. For the wave-function renormalization we find
\begin{equation*}
z_{2}=\tfrac{\varepsilon^2}{450}+O(\varepsilon^{3}) \qquad\qquad z_{4}=-\tfrac{\varepsilon^2}{900}+O(\varepsilon^{3})
\end{equation*}
\begin{equation*}
z_{6}=\tfrac{\varepsilon^2}{90}+O(\varepsilon^{3})\,, 
\end{equation*}
with all other couplings of order at least $O(\varepsilon^{3})$.
As expected the spectrum regarding $V$ agrees with \eqref{specLPATri}.
The spectrum regarding $Z$ is instead,
\begin{equation*}
\theta_2^{Z}=1-\tfrac{4}{5}\varepsilon \quad\qquad \theta_4^{Z}=2  \quad\qquad \theta_6^{Z}= 3+4\varepsilon	
\end{equation*}
\begin{equation*}
\theta_8^{Z}=4+\tfrac{64}{5}\varepsilon \quad\qquad \theta_{10}^{Z} =5+28\varepsilon \,,
\end{equation*}
which also matches \eqref{thetaZ}. This is a non-trivial fact which is not guaranteed to happen in the derivative expansion \cite{ODwyer:2007brp}.
\\

\paragraph*{\tt multi-critical}

The fact that we do not reproduce the correct anomalous dimension of the {\tt Tricritical} theory -- we have also extended the analysis to the {\tt Tetracritical} and higher {\tt multi-criticality} classes finding that the mismatch persists\footnote{We have also checked that exactly the same mismatches are found in the $m \to \infty$ limit of the PTRG--DE2 \cite{DEvsEE}.} -- signals that the DE2 we are using is intrinsically not $\varepsilon$-correct and that some new ingredient must be added. On the bright side, the $\theta^Z_i$ spectrum at order $O(\varepsilon^2)$ is instead correct, which is a non-trivial fact in its own right, shared neither by the NPRG nor by the Polchinski RG \cite{ODwyer:2007brp}.
\\

We side with \cite{ODwyer:2007brp} and believe that $\varepsilon$-correctness is an important property any RG flow should respect, while it is possible that DE4 contributions may correct the {\tt Tricritical} anomalous dimension, they surely cannot do so for higher multi-criticalities. We speculate that the solution lies in carefully considering normal ordering of operators as pointed out in the early days of the RG by Wegner \cite{Wegner:1974ph} and revised in the context of Polchinski RG by \cite{ODwyer:2007brp}. 
Work in this direction is under way \cite{:FRG:}.

\section*{Comparison with other FRG's}\label{comparison}\vspace{-0.2cm}

In this section we compare FDR with two widely used  functional RGs: non-perturbative RG (NPRG) (for a review see \cite{Berges:2000ew,Dupuis:2020fhh}) and proper-time RG (PTRG) (see \cite{Liao:1994fp}).
We begin with the LPA and then move to the full DE2. 
In particular, we show that the $m\to\infty$ limit of the PTRG reproduces the FDR beta functional for the potential but -- crucially -- differs at order DE2, confirming that FDR is a distinct approach.
Finally, we introduce the process of {\it distillation}, by which one can derive the (one-loop) FDR flow from those of the NPRG or PTRG.

\subsection*{\color{teal}Non-Perturbative RG}\vspace{-0.2cm}

The NPRG-LPA is very well studied in the literature; with the Litim regulator $R_\mu(q^2) = (\mu^2-q^2)\theta(\mu^2-q^2)$ it takes the following form 
\begin{equation}
\beta_V^{\rm NPRG} = c_d \frac{\mu^{d+2}}{\mu^2+V''}\,,
\end{equation}
where $c_{d}^{-1}\equiv(4\pi)^{\frac{d}{2}}(\frac{d}{2})!$.
The Litim regulator is PMS-optimal and gives the best LPA-NPRG estimates for the critical exponents $\nu$ and $\omega$.
As already discussed in \cite{Codello:2017hhh}, connection with perturbation theory can be made by expanding $\beta_V^{\rm NPRG}$ in powers of the cutoff:
\begin{eqnarray}
\beta_V^{\rm NPRG}&=&\mu^d c_d -  \mu^{d-2} c_dV''+ \mu^{d-4}c_d(V'')^2  \label{NPRG-LPA}\\&& 
- \mu^{d-6} c_d(V'')^3+\mu^{d-8}c_d(V'')^4
+...\nonumber\,.
\end{eqnarray}
The coefficient of $\mu^{d-d_{c}}$, when evaluated at $d=d_{c}$, is exactly the universal $\beta^{\rm DR}(d_c)$.
This is the basic property of all one-loop improved RG flows.
It is illuminating to recast \eqref{NPRG-LPA} in a form similar to our master formula \eqref{master} (together with \eqref{betaVDR}) for $\beta_V^{\rm FDR}$ by expressing $c_d$ explicitly
\begin{equation}\label{betaVLPA}
\beta_V^{\rm NPRG} (d)= \sum_{n=0}^{\infty} \mu^{d-2n} \frac{(-1)^n}{(4\pi)^{\frac{d}{2}} (\frac{d}{2})!}(V'')^n\,.
\end{equation}
We see that the only difference between the two is the dimension at which the coefficients of $(V'')^n$ are evaluated: in the NPRG-LPA they are evaluated at the fixed dimension $d$, while in the FDR-LPA they are evaluated at the critical dimensions $d_c=2n$.
In the first case -- NPRG -- the coefficients can be brought out of the sum and the resulting series is geometric, in the second case -- FDR -- they are the Taylor coefficients of an exponential. 

Repeating the analysis for a general cutoff $R_k(q^2)=k^2 r(q^2/k^2)$ allows us to write the NPRG-LPA as a sum over $d_{c}$:
\begin{equation}\label{NPRGsumoverdc}
\beta_{V}^{\rm NPRG}(d)=\sum_{d_{c}}\mu^{d-d_{c}}f(d,d_{c})\beta_{V}^{\rm DR}(d_{c})\,,
\end{equation}
where we used $\beta_{V}^{\rm DR}(d_{c})$ as given in \eqref{betaVDR}.
The ``transfer matrix'' $f(d,d_c)$ encodes how the scheme-independent DR contributions at each critical dimension $d_c$ are weighted and mixed when the flow is evaluated at the physical dimension $d$. Its explicit form is
\begin{equation}\label{fddc}
f(d,d_{c})=\frac{c_{d}}{c_{d_{c}}}\frac{d}{2}\int_{0}^{\infty}{\rm d}y\,y^{\frac{d}{2}-1}\frac{r(y)-yr'(y)}{[y+r(y)]^{\frac{d_{c}}{2}+1}}\,.
\end{equation}
That $f(d_{c},d_{c})=1$ follows directly from the standard result on scheme independence: when $d=d_c$, the  term $\mu^{d-d_c}$ must reproduce the DR beta functional exactly, irrespective of the cutoff profile $r(y)$. For the Litim regulator the integral in \eqref{fddc} gives exactly $\tfrac{2}{d}$ so $f(d,d_c)=\frac{c_{d}}{c_{d_{c}}}$ and thus the ``transfer matrix'' trades $c_{d_c}$ -- inside $\beta_{V}^{\rm DR}(d_{c})$ -- with $c_d$.
FDR, on the other hand, corresponds to the only choice where the coefficients are set constant (and thus equal to one) $f(d,d_{c})=1,\,\forall d$ -- and in particular independent of $d$.
No admissible cutoff profile $r(y)$ can realize this limit, because the integral \eqref{fddc} is generally larger than one when $d<d_c$ and vice-versa smaller when $d>d_c$ for any smooth function $r(y)$ satisfying the standard regularity and normalization conditions.
We therefore conclude that FDR is not a special case of NPRG, even if both can be written as \eqref{NPRGsumoverdc} respecting $f(d_{c},d_{c})=1$. 

In summary, one-loop exactness requires that, when expanded in powers of the cutoff $\mu$, the threshold functions of any RG scheme must reproduce the scheme-independent DR beta functionals at each critical dimension $d_c$. Concretely, this means that the coefficient of $\mu^{d-d_c}$ in the expansion, when evaluated at $d=d_c$, must coincide exactly with $\beta^{\rm DR}(d_c)$. FDR takes this principle to its logical extreme: instead of carrying the physical dimension $d$ inside the threshold functions and only matching the DR results at the special points $d=d_c$, FDR evaluates the threshold functions directly at $d_c$ from the outset. In this way, the $d$-dependence is removed from the internal structure of the beta functionals and appears only in the external cutoff factors $\mu^{d-d_c}$, which are then summed over the set of critical dimensions.

\subsection*{\color{teal}Proper-Time RG}\vspace{-0.2cm}

The general PTRG-LPA depends on the parameter $m$ and takes the following form \cite{Bonanno:2000yp,Mazza:2001bp}
\begin{equation}
\beta_{V}^{\rm PTRG} = \mu^{d}\frac{\Gamma(m-\frac{d}{2})}{(4\pi)^{\frac{d}{2}}\Gamma(m)}m^{\frac{d}{2}}\left(\frac{\mu^{2}}{\mu^{2}+\tfrac{1}{m}V''}\right)^{m-\frac{d}{2}}\,.
\end{equation}
For $m=1$ it corresponds to the Callan-Symanzik RG (CSRG) flow, while for $m=\tfrac{d}{2}+1$ (i.e.\ $m=\tfrac{5}{2}$ in $d=3$) it reproduces the optimal Litim NPRG-LPA analysed above \cite{Litim:2001hk}. It also includes the Wegner-Houghton RG \cite{Wegner:1972ih} in the case $m=\tfrac{3}{2}$, so this comparison covers a great part of the functional RG LPA history.
More importantly for our discussion, in the $m\to\infty$ limit we recover our FDR-LPA beta functional \eqref{betaVL1}, up to a trivial field redefinition that moves the $4\pi$ factors from the $d$-dependent pre-factor to the $d$-independent exponential,
\begin{equation}\label{PTRGLPA}
\lim_{m\to \infty }\beta_{V}^{\rm PTRG}=\frac{\mu^d}{(4\pi)^{\frac{d}{2}}} e^{-\frac{V''}{\mu^2}}\,.
\end{equation}
Thus -- {\it at LPA level} -- FDR is equivalent to the PTRG in the limit $m\to\infty$.
Finally, one may argue that the LPA equivalence between PTRG and FDR could ``explain'' the quality of the PTRG results for critical exponents, which has been a puzzle and a source of discussion \cite{Litim:2001ky,Litim:2002xm,Litim:2001hk}.

The PTRG-DE2 in the limit $m\to\infty$ (we consider here the ``spectrally adjusted'' case) reads \cite{Mazza:2001bp}:\\
\begin{widetext}
\begin{eqnarray}\label{PTRGbetaZ}
\beta_Z &=&
\left\{
-\frac{1}{6}\frac{\mu^{d-6}}{(4\pi)^{\frac{d}{2}}} \frac{(V''')^2}{(1+Z)^2}
-\frac{\mu^{d-2}}{(4\pi)^{\frac{d}{2}}} \frac{Z''}{1+Z}
+\frac{10-d}{6}\frac{\mu^{d-4}}{(4\pi)^{\frac{d}{2}}} \frac{V'''Z'}{(1+Z)^2}
-\frac{d^{2}-18d-4}{24}\frac{\mu^{d-2}}{(4\pi)^{\frac{d}{2}}}\frac{(Z')^2}{(1+Z)^2}
\right\} e^{ -\tfrac{V''}{\mu^2 (1+Z)}}\,. \nonumber\\
\end{eqnarray}
\end{widetext} 
%
Comparing the FDR-DE2 beta functional \eqref{betaZ} with its PTRG counterpart \eqref{PTRGbetaZ} reveals both structural similarities and key differences. Both expressions share the same overall architecture: a sum of terms involving $V'''$, $Z'$, $Z''$, and their combinations, all multiplied by an exponential factor that encodes the threshold behavior. Moreover, in both cases the terms are organized as an expansion in powers of the cutoff $\mu$, with the leading contributions to the anomalous dimension arising from $\mu^{d-6}(V''')^2$ and $\mu^{d-2}Z''$.
The differences, however, are equally telling. In the PTRG expression, the coefficients of the various terms are polynomials in $d$ inherited from the integration over the proper-time parameter, and the $4\pi$ factors carry a $d$-dependent exponent. In FDR, by contrast, the coefficients are pure numbers -- independent of $d$ -- and the $4\pi$ factors are fixed at their dimensional values at which they first appear, as they arise from evaluating the DR contributions at the critical dimensions $d_c$. 
%
%
Despite these differences, the close structural resemblance explains why PTRG and FDR yield comparable results at the level of critical exponents, while the $d$-independent nature of the FDR coefficients is ultimately responsible for its improved convergence properties -- as can be seen, for example, in Figure~\ref{EtaFDRDE2} in the case of the anomalous dimension. 

\subsection*{\color{teal}{\it Distillation} of Functional RG Flows}\vspace{-0.2cm}

We call {\it distillation} the process by which the FDR beta functions or functionals can be derived from the corresponding ones in any other one-loop exact RG scheme, such as NPRG or PTRG. Starting from the dimension-full beta functional in one of these schemes, one first extracts all scheme-independent terms (i.e., the DR beta functionals) -- either by employing a cutoff such as the one proposed in \cite{Baldazzi:2020vxk}, or by simply selecting all terms proportional to a given power of $\mu^{d-d_c}$ and evaluating them at $d=d_c$ (see also the appendix of \cite{Codello:2017hhh}). These terms are then inserted into our master formula \eqref{masterformula}, where the sum over the specific set of $d_c$'s is performed to return the FDR beta functionals.

We will use $\beta_Z$ at LPA level in the NPRG as an example to illustrate the procedure.
With the Litim regulator it takes the form
\begin{equation}
\beta_Z^{\rm NPRG}=-c_d\,\mu^{d+2}\frac{(V''')^2}{(\mu^2+V'')^4}\,.
\end{equation}
Expanding the denominator and collecting powers of the cutoff gives
\begin{widetext}
\begin{eqnarray}\label{BetaZLPAexp}
\beta_Z^{\rm NPRG}&\!\!=\!\!&-c_d\Big\{\mu^{d-6}-4\mu^{d-8}V''+10\mu^{d-10}(V'')^2
-20\mu^{d-12}(V'')^3+35\mu^{d-14}(V'')^4+...\Big\}(V''')^2\,.
\end{eqnarray}
\end{widetext}
We see that the leading term starts with $d_c=6$ as expected.
The DR beta functionals are easily extracted and can be conveniently written as
\begin{equation}\label{c2n4}
\beta_Z^{\rm DR} (d_c = 2n+4) = 
\frac{(-1)^n  (V'')^{n-1}}{6(4\pi)^{n+2}(n-1)!} (V''')^2\,,
\end{equation}
with $n=1,2,3,...$ corresponding to $d_c=6,8,10,12,...$\,.
Inserting into the master formula \eqref{masterZ} and summing over the above set of upper critical dimensions
\begin{eqnarray*}
\beta^{\rm FDR}_Z&=&(V''')^2 \sum_{n=1}^{\infty} \mu^{d-(2n+4)} \frac{(-1)^n(V'')^{n-1}}{6(4 \pi)^{n+2}(n-1)!} \\
&=&-\frac{(V''')^2}{6(4\pi)^3} \mu^{d-6} \sum_{n=1}^{\infty}  \frac{1}{(n-1)!} \left(\frac{-V''}{4\pi\mu^2} \right)^{n-1}\\
&=&   -\mu^{d-6}\frac{(V''')^2}{6(4\pi)^3} e^{ -\frac{V''}{4 \pi \mu^2}} \,.
\end{eqnarray*}    
gives back the anomalous dimension functional \eqref{betaZLPA}, as expected.

In practice the  distillation procedure may be the fastest way to obtain the FDR beta functionals, if the corresponding functional RG results are readily available in the literature; and,  in any case, they may serve as useful double-checks of computations done directly using DR.
Finally, we note that the distillation process does not preserve exactness beyond one-loop, as it can also be applied to non-exact flows such as the PTRG.
The only requirement for the application of the procedure is that the target flow be one-loop correct.

\section*{Conclusion and Outlook}\vspace{-0.2cm}

In this paper, we have demonstrated that {\it functional dimensional regularization} (FDR), as introduced in \cite{Beretta:2026zcy}, exhibits all the expected properties of a functional renormalization group (RG) flow and establishes itself as a competitive new framework. Its performance with respect to the rate of convergence and the stability of the local potential approximation (LPA) and the derivative expansion to second order (DE2) compares favorably to existing methods. FDR also delivers high-quality estimates for critical properties -- most notably the anomalous dimension in $d=3$ -- and, remarkably, it even proves capable of correctly recovering the results from the $\varepsilon$-expansion for the {\tt Ising} universality class. Furthermore, we presented the characteristic spike plot and eigen-perturbation analysis, and we closed with a detailed comparison to other functional RG approaches. This comparison motivated the introduction of the {\it distillation} procedure, which provides a systematic way to obtain the FDR flow from any other known functional flow.

Still, many open questions remain. On the one hand, we have observed that at the LPA level the FDR equations are equivalent to those of the proper-time RG, while at second order in the derivative expansion the functional forms are very similar, differing only in the coefficients -- which are $d$-dependent in the proper-time RG case, but $d$-independent in the FDR case. This structural similarity may explain the ``unreasonable success'' of early proper-time estimates for critical exponents \cite{Bonanno:2000yp}. On the other hand, it has been shown that the proper-time RG is not an exact flow \cite{Litim:2001hk}, so the question arises whether the same arguments can be applied to the one-loop-improved FDR flow studied in this paper. If so, the systematic extension of the FDR approach beyond one loop must proceed along what we called options (a) and (b): to incorporate higher-loop contributions---along with all the new critical dimensions they introduce -- and to extend the range of summation in our master formula \eqref{masterformula}. Another crucial point is to preserve $\varepsilon$-correctness at all multi-critical fixed points \cite{ODwyer:2007brp,:FRG:}, which, as we have seen, fails at DE2 beyond the \texttt{Ising} universality class. We think that all of these options, when properly explored, will give us clues that can show us the way toward a complete development of the FDR approach.

The results of this paper motivate us to further test our DE2 flow equations in $d=2$ and across the range $2\leq d \leq4$, as well as to include higher-dimensional operators such as those that appear at next order,  DE4. Further benchmarks will come by applying FDR directly in other contexts, such as $O(N)$-models \cite{Beretta:2026avu} and multi-field theories with more general symmetries. Another promising direction is to tackle open questions in the realm of odd models, including the non-unitary {\tt Lee-Yang} universality class \cite{Zambelli:2016cbw} and its multi-critical cousins \cite{Codello:2017epp}. Further interesting possibilities include combining FDR with the ideas of essential RG \cite{Baldazzi:2021ydj}; introducing fermionic degrees of freedom to study scalar-Yukawa theories \cite{Vacca:2015nta,Jack:2024sjr}; and exploring higher-derivative theories \cite{Safari:2017irw}. Applications beyond the statistical physics are equally compelling -- for instance, investigating the implications of FDR on the RG flow and the stability of the Standard Model. Finally, it is of considerable interest to study the implications of this new framework for gauge theories (see the recent first application \cite{Kluth:2026djl}) and gravity, where non-orthodox uses of dimensional regularization have already partially paved the way -- although not yet incorporating IR critical dimensions (see \cite{Kluth:2024lar} and references therein). All these directions will be the subject of future work.

\subsection*{\color{teal}Acknowledgments}\vspace{-0.2cm}

We thank D. Zappal\`a, G. P. Vacca, and K. Falls for valuable comments and feedback.
The authors gratefully acknowledge financial and computational support from the CSIC grant I+D-2022-22520220100174UD.


\bibliography{referencesFDR}

\end{document}